\documentclass[pra,twocolumn,superscriptaddress,floatfix,longbibliography]{revtex4-2}

\usepackage{amsfonts}
\usepackage{amsmath}
\usepackage{graphicx}
\usepackage{amssymb}
\usepackage{color}
\usepackage{subfigure}
\usepackage{mathtools}
\usepackage{tikz-cd}
\usepackage[colorlinks]{hyperref}
\usepackage[normalem]{ulem}

\usepackage{bbm}

\newcommand{\normord}[1]{\mathopen{:}#1\mathclose{:}}

\begin{document}
\title{Corrections to the Optomechanical Hamiltonian from Quadratic Fluctuations of a Moving Mirror}

\author{Salvatore Butera}
\affiliation{School of Physics and Astronomy, University of Glasgow, Glasgow G12 8QQ, United Kingdom}

\begin{abstract}
We extend the theory of the radiation pressure to include quadratic fluctuations in the position of a moving mirror. This enables the introduction of a generalized radiation pressure operator that captures higher-order effects in the mirror-field coupling. For mechanical resonators with frequencies comparable to the fundamental cavity frequency, the resulting corrections to the optomechanical Hamiltonian are quadratic in the standard radiation pressure coupling. These corrections are expected to play a significant role in the strong coupling regime of optomechanics, electromechanics, and superconducting circuit analogs.
\end{abstract}

\maketitle

\section{Introduction}
Optomechanics is the rapidly developing field of research that studies fundamental principles and practical applications of the interplay between light and mechanical motion~\cite{Aspelmeyer_RMP}. This interaction originates from the exchange of momentum between light and macroscopic objects, giving rise to a radiation force commonly referred to as \emph{radiation pressure}. Radiation pressure underpins numerous phenomena across diverse scales, from the deflection of comet tails by the solar wind~\cite{Comet1951}, to the precise manipulation~\cite{Ashkin1978,Ashkin1986,Ashkin1987} and cooling of atoms, molecules~\cite{LaserCooling,chan2011laser} and nanoparticles~\cite{Aspelmeyer-Science-2020} using light. Over the past few decades a surge of activity has focused on this topic, partly motivated by the essential role that radiation forces play in the physics of optical interferometers. In gravitational wave detectors~\cite{Abramovici-science-1992,virgo-2012}, for instance, fluctuations in the optical field act as a noise heating up mechanical modes and thus limiting the resolution by which the position of a moving mass can be measured~\cite{Caves-PRL-1980,Loudon-PRL-1981,Vignes-1985,Walls-PRA-1993}. Recently, pioneering back-action cooling of mechanical motion to the quantum ground state was achieved~\cite{OConnell-Science-2010,chan2011laser,teufel2011sideband, Painter2012} even with macroscopic kilogram-scale oscillators~\cite{abbott2009observation}.

The recent experimental progress in preparing and tailoring quantum states of light, together with the developments in nano-fabrication technologies and cooling techniques, has advanced optomechanics into the quantum domain~\cite{Paternostro_Review}. These advances have enabled numerous applications across diverse research areas, including high-precision metrology and sensing~\cite{Stange-PhysToday-2021}, quantum information storage~\cite{MechQbit-PRX-2021,navarathna2022good,Yang2024} and processing~\cite{Rabl2012}, quantum thermodynamics and thermal machines that use radiation pressure as a mechanism to exchange energy at the level of single excitations~\cite{Nori-QuantumHEatEng-2007,Bariani-QuantumHEatEng-I,Bariani-QuantumHEatEng-II,Bariani-QuantumHEatEng-III,Nori-QuantumHEatEng-2023}.

The quantum theory of the radiation pressure was developed in the 90's of the past century by C.K.~Law, whose pioneering work laid the theoretical foundation for modern quantum optomechanics~\cite{Law1995}. Notably, the radiation pressure interaction is inherently nonlinear due to its dependence on optical intensity and the feedback loop generated by the field-mirror backreaction process: fluctuations in the electromagnetic field affect the position of a moving mirror, which in turn influences the field dynamics. In more technical terms, through the quantization procedure field operators are explicitly defined in terms of the geometry of the system and, ultimately, the position of the mirror. To handle this complexity, it is standard practice in optomechanics to work in the limit of small displacements of the mirror from its equilibrium position within a confining potential~\cite{Law1995,Aspelmeyer_RMP}. This assumption allows linearization of the Hamiltonian with respect to mechanical displacements, thereby fixing the quantization geometry for the field and the associated quantum operators. Such an approximation is justified as long as the fluctuations in the mirror's position are much smaller than the characteristic wavelengths of the field modes interacting with the mirror, and provided the system operates in the weak coupling regime.

This approximation is challenged by the ongoing miniaturization of optomechanical devices, which has pushed experiments toward the single-photon strong coupling regime. This regime occurs when the coupling rate of a single photon is of the order of, or exceeds the loss rates of the mechanical or optical resonators~\cite{kippenberg2008cavity,Cleland2010,Girvin2011_SinglePhoton}. Such a condition can be experimentally achieved using hybrid electromechanical systems~\cite{Regal2008,lahaye2009nanomechanical,teufel2011circuit,UltraStrong_Review}, where the strength of the radiation pressure can be enhanced by several orders of magnitude by introducing superconducting qubits as mediators of the interaction between the mechanical oscillator and a microwave cavity~\cite{Rouxinol-2016,Sillanpaa2014,pirkkalainen2015,Vitali2021,Hakonen-PRB-2022}.
Alternatively, superconducting electronic circuits can provide an analog platform to reach this regime. In fact, a resonator magnetically coupled to a microwave coplanar waveguide through a superconducting quantum interference device (SQUID) can be used to simulate a high-frequency vibrating mirror~\cite{Johansson-PRL-2009,Johansson-PRA-2010,Wilson-DCE-Analog-2011,Butera-PRA-2019}.

In the strong coupling regime, the nonlinear nature of the radiation pressure leads to effects that depart from classical behaviour and facilitates the production of nonclassical states of light and mechanical motion~\cite{Knight1997,Tombesi1997,Savasta2015}.
Nonlinear effects have also been demonstrated to play a pivotal role, both in dressing the bare optomechanical vacuum with virtual excitations~\cite{Giulio-PRL-2013,Armata-PRD-2015,Armata-PRD-2017}, and in determining the back-action dynamics induced by the dynamical Casimir emission on a moving mirror~\cite{KardarRMP1999}, both in terms of quantum friction~\cite{KardarRMP1999,Savasta-PRX-2018,Butera-PRA-2019,Savasta-PRL-2019} and vacuum induced dephasing of the mechanical oscillations~\cite{Butera-EPL-2019,butera2023noise,Dalvit-PRL-2000,MaiaNeto-PRA-2000}.

The ability to probe nonlinear effects of radiation pressure raises questions about the validity of the linear displacement approximation in the optomechanical Hamiltonian, thereby motivating further investigations to account for higher-order effects of the mirror fluctuations~\cite{Tufarellu2018}. 
Building upon Law's original work~\cite{Law1995}, this paper takes a step in this direction by developing a complete and self-consistent extension of the quantum theory of radiation pressure that includes the effects of quadratic mechanical fluctuations. This study is driven by the growing interest in quantum phenomena arising from higher-order optomechanical coupling. For instance, the quadratic coupling discussed in this paper has been proposed as a means for realizing quantum nondemolition measurements of phonon number~\cite{Thompson2008,Miao2009,Marquardt2012}, measurement of phonon shot noise~\cite{Clerk2010}, as well as cooling and squeezing of mechanical motion~\cite{Meystre2008,Girvin2010,Vitali2014}. Optomechanical setups have been specifically engineered to investigate the effects of this higher-order interaction, including membranes embedded in standard Fabry-P\`erot cavities~\cite{Thompson2008}, microdisk-cantilever systems~\cite{Davis2014}, microsphere-nanostring systems~\cite{Bowen2016}, paddle nanocavities~\cite{Barclay2015}, and photonic crystal coupled to micro-beams~\cite{Painter2015}.

In this paper, we focus, for simplicity, on the one-dimensional system composed of an optical cavity terminated on one side by a dynamic mirror. In this configuration, the electromagnetic vector potential is equivalent to a massless scalar field.
We show that nonlinear fluctuations of the mirror give rise to higher-order mirror-field interaction processes, which allow us to define a generalized radiation pressure operator. In addition to reproducing the standard radiation pressure, which describes the interaction between pairs of bare photons and the moving mirror, this operator also includes corrections due to the mixing of optical modes induced by mechanical motion.
By performing an order-of-magnitude analysis we demonstrate that, for high-frequency mechanical oscillators with frequencies comparable to the fundamental cavity frequency~\cite{Cleland2010,primo2023dissipative}, these corrections are quadratic in the optomechanical single-photon interaction strength. Their effects are thus expected to be significant in processes that are quadratic in the mirror-field interaction, yet they have been largely overlooked within the literature.

The paper is organized as follows: In Sec.~\ref{Sec:QRP}, we briefly review the quantization procedure for a scalar field constrained by a massive and dynamic boundary condition, introducing the notation that will be used within the paper. We report the main results of this work in Sec.~\ref{Sec:SD}. Specifically, in Sec.~\ref{Sec:ampl}, the optomechanical Hamiltonian expanded up to quadratic fluctuations of the mirror's position is presented, and the generalized radiation pressure operator introduced. Rather expressing the Hamiltonian ab initio in terms of mirror-position-dependent annihilation and creation operators for the field modes, we find it convenient to develop the theory by employing modes amplitudes and their conjugate momenta as the primary dynamical variables. The expanded optomechanical Hamiltonian, expressed in the standard ladder operators formalism, is then presented in Sec.~\ref{Sec:ladder}. Finally, we discuss these results and draw our conclusions in Sec.~\ref{Sec:Conclusions}. For completeness, and to connect with Law's original approach, the optomechanical Hamiltonian expressed in terms of position-dependent annihilation and creation operators for the field modes is given in Appendix~\ref{app:A}, together with the quadratic expansion of these operators. The key steps in the derivation of the expanded Hamiltonian is reported in the Appendix~\ref{app:B}, for the convenience of the reader.

\section{Quantization procedure\label{Sec:QRP}}
\begin{figure}[t]
    \centering
    \includegraphics[width = 0.45 \textwidth]{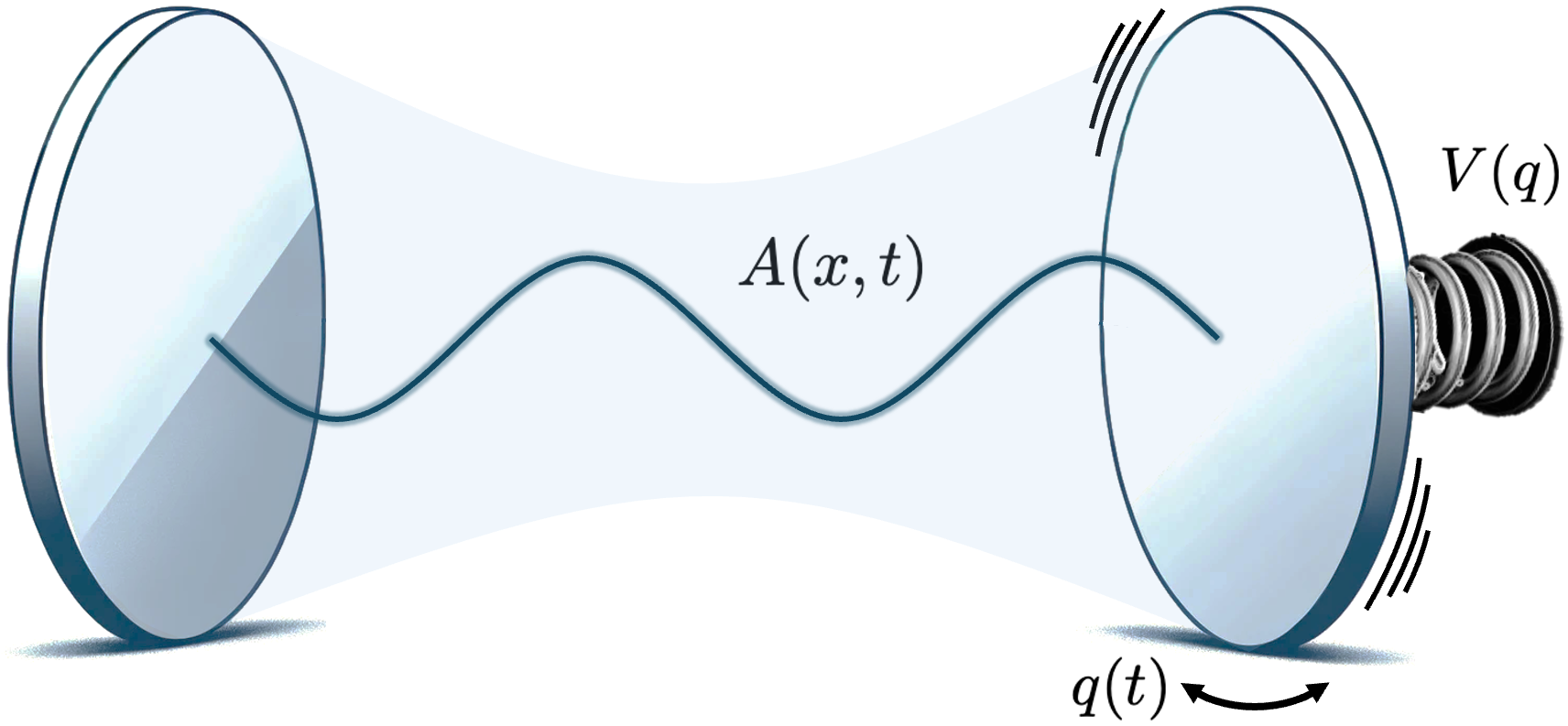}
    \caption{Schematic representation of an optical cavity bounded by a dynamic mirror at position $q(t)$. The mirror is confined by a potential $V(q)$, and interacts via radiation pressure with the scalar field $A(x)$ enclosed within the cavity.}
\label{Fig1}
\end{figure}

Let us consider a one-dimensional optical cavity bounded by a fixed mirror at position $x = 0$ and a moving mirror of mass $m$ at $ x = q(t) $, where $ q(t) $ denotes its time-dependent position. The moving mirror is confined by a harmonic potential of the form $ V(q) \equiv m\Omega^2 (q-d)^2/2$, where $\Omega$ is the mirror’s bare mechanical oscillation frequency, and $d$ is the cavity length corresponding to the mirror's equilibrium position at the potential minimum. Both mirrors are assumed to be perfect reflectors for a massless scalar field $ A(x, t) $ enclosed within the cavity, thus requiring the field to vanish at their positions. This leads to the following boundary conditions for the field: $ A(0, t) = A(q(t), t) = 0 $. A schematic of the system at hand is shown in Fig.~\ref{Fig1}.

By modeling the moving mirror as a non-relativistic massive object, the dynamical properties of the system are described by the following Lagrangian $(c=1)$:
\begin{equation}
	L = \frac{1}{2}m\dot{q}^2(t)-V(q) + \frac{1}{2} \int_0^{q(t)} dx \,\big[(\partial_t A)^2 - (\partial_x A)^2\big].\label{L}
\end{equation}
For ease of notation, we use overdots to indicate total time derivatives: $\dot{q}\equiv dq/dt$. Let us expand now the field in terms of the complete set of eigenmodes that instantaneously satisfy the boundary conditions. Specifically, we write: $A(x,t) = \sum_k Q_k(t) \varphi_k[x,q(t)]$ $(k \in \mathbb{N})$, with the eigenmodes having the form:
\begin{equation}
	\varphi_k[x,q(t)] \equiv \sqrt{\frac{2}{q(t)}}\sin\left\{\omega_k[q(t)]\,x\right\},\label{phi_k}
\end{equation}
given the frequencies $\omega_k[q(t)] \equiv k \pi/q(t)$. These eigenmodes implicitly depend on time through the mirror's position $q(t)$. By using the field expansion given above, with the eigenmodes defined in Eq.~\eqref{phi_k}, and carrying out the spatial integrals, the Lagrangian in Eq.~\eqref{L} can be recast in the form~\cite{Law1995}:
\begin{multline}
	L = \frac{1}{2}m\dot{q}^2-V(q) +\frac{1}{2}\sum_k\left(\dot{Q}_k^2-\omega_k^2(q)Q_k^2\right)\\
	+\frac{\dot{q}}{q}\sum_{k}\dot{Q}_k Q_k^{(1)} - \frac{\dot{q}^2}{2q^2}\sum_{k} Q_k Q_k^{(2)}.\label{L2}
\end{multline}
Notice that we dropped time-dependence from variables for brevity and ease of notation. In Eq.~\eqref{L2}, we defined $Q_k^{(1)}$, $Q_k^{(2)}$ according to the recursive relation:
\begin{align}
	Q_k^{(0)} &\equiv Q_k, \label{Q0}\\
	Q_k^{(n+1)} &\equiv \sum_j g_{jk} Q_j^{(n)} \qquad (n \in \mathbb{N}),\label{Qn}
\end{align}
with the coefficients:
\begin{align}
	g_{jk} &\equiv q \int_0^q dx\left(\varphi_k\frac{\partial\varphi_j}{\partial q}\right) \nonumber\\
	& = \left\{
        \begin{array}{cc}
                (-1)^{k+j}\left(\frac{2 k j}{ k^2- j^2}\right) & \quad (k\neq j) \\
                0 & \quad (k = j)
    	\end{array}.
    \right.\label{g_kj}
\end{align}
We remark that, to obtain Eq.~\eqref{L2}, we also made use of the completeness relation: $\sum_s \varphi_s[x,q(t)]\varphi_s[y,q(t)] = \delta(x-y)$ to write:
\begin{equation}
\sum_s g_{ks}g_{js} = q^2 \int_0^q dx \, \frac{\partial\varphi_k(x)}{\partial q} \frac{\partial\varphi_j(x)}{\partial q}.\label{compl}
\end{equation}
The dimensionless $g_{jk}$ coefficients introduced in Eq.~\eqref{g_kj} quantify the overlap between the $j$ and $k$ field's modes, induced by the displacement of the mirror. As such, the quantity $Q_k^{(n)}$ can be interpreted as the fluctuation amplitude in the $k-$th mode, which arises due to the mixing of the $k$ mode with the other field modes, generated by mechanical fluctuations of order $n$.

The Lagrangian in Eq.\eqref{L2} fully describes the coupled mirror-field dynamics. In addition to the intrinsic dependence of field mode frequencies on the mirror's position, mechanical motion introduces two non-adiabatic interaction terms. These contributions, represented by the last two terms in Eq.\eqref{L2}, are linear and quadratic in the mirror's velocity and involve linear $Q_k^{(1)}$ and quadratic $Q_k^{(2)}$ field's fluctuations, respectively.

Aiming to quantize the theory, we derive the Hamiltonian of the system, which is formally defined as:
\begin{equation}
H \equiv p\dot{q} + \sum_k P_k \dot{Q}_k - L.
\end{equation}
Here, 
\begin{align}
p &\equiv \frac{\partial L}{\partial \dot{q}} = m\dot{q} + \frac{1}{q}\sum_{k}P_k Q_k^{(1)},\label{p}\\
P_k &\equiv \frac{\partial L}{\partial \dot{Q}_k} = \dot{Q}_k + \frac{\dot{q}}{q} Q_k^{(1)},\label{Pk}
\end{align}
are the mirror's and field's canonical momenta, respectively. By using these definitions, the Hamiltonian can be written in the form~\cite{Law1995}:
\begin{multline}
H = \frac{\left[p+\Gamma(q)\right]^2}{2m} + V(q) + \frac{1}{2}\sum_k\left[P_k^2+\omega_k^2(q) Q_k^2\right].\label{Hq}
\end{multline}
We have thus obtained that, in the Hamiltonian formalism, the non-adiabatic mirror-field coupling is described by the operator:
\begin{equation}
\Gamma(q) \equiv - \frac{1}{q} \sum_{k}P_k Q_k^{(1)}. \label{Gamma}
\end{equation}
The theory is finally quantized by promoting dynamical variables to operators and imposing the canonical commutation relations between amplitudes and the corresponding conjugate momenta: $[\hat{q},\hat{p}] = i\hbar$ and $[\hat{Q}_k,\hat{P_j}] = i\hbar\,\delta_{kj}$. Mirror-position-dependent annihilation $\hat{a}_k(q)$ and creation $\hat{a}_k^\dag(q)$ operators for the field modes are introduced in Appendix~\ref{app:A}, where the Hamiltonian expressed in terms of these operators is given for completeness.

In the next section, we present an approximate version of the Hamiltonian, that is obtained by expanding its dependence on the mirror’s position $q(t)$ up to quadratic order in the fluctuations of the mirror around the equilibrium position within the confining potential. Rather than retracing Law's original approach, which first expresses the Hamiltonian in terms of the annihilation $\hat{a}_k(q)$ and creation $\hat{a}_k^\dag(q)$ operators for the field modes and then expands these operators with respect to mirror's fluctuations, we find it more convenient to perform the expansion directly in Eq.~\eqref{Hq}, thus treating the amplitudes $Q_k(t)$ and the corresponding momenta $P_k(t)$ as the fundamental dynamical variables. Annihilation $\hat{a}_{k0}\equiv \hat{a}_k(d)$ and creation $\hat{a}^\dag_{k0} \equiv \hat{a}_k^\dag(d)$ operators, corresponding to the cavity's equilibrium configuration are then introduced only after the expansion is complete, thereby reconnecting with the standard formalism. For completeness, and to connect with Law’s original approach, the ladder operators $\hat{a}_k(q)$ and $\hat{a}_k^\dag(q)$ expanded up to quadratic fluctuations of the mirror's position are also given in Appendix~\ref{app:A}.

\section{Quadratic approximation\label{Sec:SD}}
\subsection{Field amplitude and momentum formalism \label{Sec:ampl}}
In the previous section, we derived the exact Hamiltonian describing a perfectly reflecting, non-relativistic mirror interacting with a massless scalar field. Upon quantization, this interaction is described by the $\hat{\Gamma}(\hat{q})$ operator appearing in the mirror's kinetic energy, as well as by the dependence of the field modes' frequencies on the mirror's position operator.
The intricate dependence of the Hamiltonian on the mirror's position can be approximated by working in the assumption of small mechanical oscillations $x(t)$ of the mirror around the minimum of its confining potential. That is, we set: $q(t) = d + x(t)$, with $x(t)\ll d$. As noted in the Introduction, this assumption holds if the mechanical motion occurs over length scales much smaller than the characteristic wavelength of the optical modes interacting with the mirrors, a condition typically satisfied in optomechanics.

While the standard theory of radiation pressure interaction considers only linear mirror fluctuations~\cite{Law1995,Aspelmeyer_RMP}, the objective of this section is to extend the model by including quadratic contributions of the mirror's displacement operator $\hat{x}$.
To this end, we expand both the modes frequencies $\omega_k(\hat{q})$ and the operator $\hat{\Gamma}(\hat{q})$ up to second order in $\hat{x}$. Indicating with $\omega_{k0}\equiv \omega_k(d)$ and $\hat{\Gamma}_0\equiv \hat{\Gamma}(d)$ the values these quantities take in the geometrical configuration corresponding to the equilibrium position of the cavity,
the Hamiltonian can be approximated as:
\begin{multline}
\hat{H} \approx \frac{\hat{p}_{\rm kin}^2}{2m} + V(\hat{x}) + \frac{1}{2}\sum_k \big(\hat{P}_k^2+\omega_{k0}^2 \hat{Q}_k^2\big) \\- \frac{\hat{x}}{d} \sum_k\omega_{k0}^2 \hat{Q}_k^2 + \frac{3\hat{x}^2}{2d^2} \sum_k\omega_{k0}^2 \hat{Q}_k^2, \label{Eq:Law_H_2nd}
\end{multline}
where $\hat{p}_{\rm kin}\equiv \hat{p}+\hat{\Gamma}_0\left(1-\hat{x}/{d}\right)$ is the kinetic momentum operator of the mirror. It is convenient removing the mirror-field coupling from the mirror's kinetic energy, by performing a space-dependent unitary transformation $\hat{T}(\hat{x})$ (that is, a gauge transformation) on the Hamiltonian. This is achieved by choosing:
\begin{equation}
	\hat{T}(\hat{x}) \equiv \exp \bigg[-i \frac{\hat{x}}{\hbar}\Big(1-\frac{\hat{x}}{2d}\Big)\hat{\Gamma}_0\bigg].\label{T}
\end{equation}
Indeed, performing this transformation on $\hat{p}_{\rm kin}$ yields:
\begin{align}
	\hat{p}'_{\rm kin} \equiv \hat{T}^\dag(x)\hat{p}_{\rm kin}\hat{T}(x) = \hat{p},
\end{align}
so that the rotated kinetic momentum operator of the mirror reduces to its canonical momentum operator $\hat{p}$. The field modes amplitudes $\hat{Q}_k$ transform instead as:
\begin{align}
	\hat{Q}_k' &\equiv \hat{T}^\dag \hat{Q}_k \hat{T} \nonumber\\
	&= \hat{Q}_k - \frac{\hat{x}}{d}\left(1-\frac{\hat{x}}{2d}\right)\hat{Q}_k^{(1)} + \frac{\hat{x}^2}{2d^2} \hat{Q}_k^{(2)},\label{Q'}
\end{align}
and an analogous transformation applies to the field's conjugate momenta $\hat{P}_k$ (the definitions of the $n$-th order momenta $\hat{P}_k^{(n)}$ follows those of the amplitudes in Eqs.~\eqref{Q0} and \eqref{Qn}). Note that Eq.~\eqref{Q'}, together with the corresponding transformation for $\hat{P}_k$, carries the same information as would be obtained by expanding the position-dependent annihilation and creation operators according to Law's original approach (see Appendix~\ref{app:A}).
Effectively, the gauge transformation gives the mirror-field coupling the form of a potential energy relating both the mirror's and field's oscillation amplitudes. This can be deduced from the transformed Hamiltonian $\hat{H}' \equiv \hat{T}^\dag(\hat{x}) \hat{H} \hat{T}(\hat{x})$ that, after some algebra (see Appendix~\ref{app:B}), can be cast in the form:
\begin{equation}
	\hat{H}'= \hat{H}_{m}+ \hat{H}_{f} + \hat{H}_{\rm int},\label{H'}
\end{equation}
where
\begin{align}
	\hat{H}_{m} &= \frac{\hat{p}^2}{2m} + V(\hat{x}), \label{Hm}\\
	\hat{H}_{f} &=\frac{1}{2}\sum_k \big(\hat{P}_k^2+\omega_{k0}^2 \hat{Q}_k^2\big),
\end{align}
are the Hamiltonians that describe the free evolution of the mirror and the field, respectively, in the configuration corresponding to the equilibrium length $d$ of the cavity, while
\begin{equation}
	\hat{H}_{\rm int} = -\frac{\hat{x}}{d} \hat{F}_0 + \frac{\hat{x}^2}{d^2} \bigg(\frac{3}{2}\hat{F}_0+\hat{F}_1\bigg)\label{Hint}
\end{equation}
is the sought generalized version of the radiation-pressure coupling, up to quadratic order in the mirror fluctuations. Here:
\begin{align}
	\hat{F}_0&\equiv \sum_k \omega_k^2 \hat{Q}_k\left(\hat{Q}_k+\hat{Q}_k^{(1)}\right)\nonumber\\
	&=\sum_{k,j}\left(-1\right)^{k+j} \omega_{k0} \omega_{j0} \hat{Q}_k \hat{Q}_j,\label{F0}
\end{align}
is (up to a vacuum energy term) the standard radiation pressure generated by pairs of bare photons scattered by the mirror's vibration, while
\begin{align}
	\hat{F}_1&\equiv \sum_k \frac{\omega_k^2}{2} \left[\hat{Q}_k\left(\hat{Q}_k^{(1)}+\hat{Q}_k^{(2)}\right)+\hat{Q}_k^{(1)}\left(\hat{Q}_k+\hat{Q}_k^{(1)}\right)\right]\nonumber\\
	&=\sum_{k,j}\left(-1\right)^{k+j} \omega_{k0} \omega_{j0} \hat{Q}_k \hat{Q}_j^{(1)},\label{F1}
\end{align}
is a new result, capturing a higher-order contribution to the radiation pressure. Specifically, it describes the scattering of couples of photons due to the mechanical fluctuations, where one of the interacting photons results from the mechanically induced modes mixing [see discussion after Eq.~\eqref{compl}]. Eqs.~\eqref{F0} and \eqref{F1} can be unified by introducing the generalized radiation-pressure operator (having dimensions of an energy):
\begin{equation}
	\hat{F}_n\equiv \sum_{k,j}\left(-1\right)^{k+j} \omega_{k0} \omega_{j0} \hat{Q}_k \hat{Q}_j^{(n)},\label{Fn}
\end{equation}
with $n=0,1$ within the quadratic approximation here considered. The combined Eqs.~\eqref{Hint} and \eqref{Fn} generalize the standard optomechanical interaction, and represent the main result of this work. In fact, in Eq.~\eqref{Hint} we recognize the well-known linear radiation pressure interaction $\sim \hat{x} \hat{F}_0$~\cite{Law1995}, and have further obtained the sought quadratic corrections $\sim \hat{x}^2\hat{F}_{0},\, \hat{x}^2\hat{F}_{1}$. Given the harmonic potential $V(x) = m\Omega^2 x^2/2$ trapping  the mirror, these corrections can be physically interpreted as generating a dynamic optical spring effect shifting the mechanical oscillation frequency $\Omega$.
Notice that both $\hat{F}_0$ and $\hat{F}_1$ are affected by ultraviolet divergences that cannot be cured with standard renormalization techniques~\cite{milonni-book}. This can be inferred by careful inspection of Eq.~\eqref{F0} and \eqref{F1}. Given the strength of the vacuum fluctuations of the modes amplitudes: $\langle\hat{Q}_k^2\rangle = Q_{k,{\rm zpf}}^2 \equiv \hbar/(2\omega_{k0})$, we deduce that the coupling strength between the mirror and a pair of $k,s$ photons (in the limit $k,s\to\infty$) is proportional to $({\omega_k\omega_s})^{1/2}$ and $\omega_k\omega_s$, in the former and latter case, respectively. Similarly to Bethe's original derivation of the Lamb shift in the hydrogen atom~\cite{Bethe-Lamb-1947,Ashcroft-book}, such a patological  behaviour is likely a consequence of modeling the mirror as a non-relativistic object.
Nevertheless, this approximation remains valid for energies significantly smaller than the mirror's rest energy $mc^2$. In this respect, note that materials become transparent to electromagnetic radiation at frequencies of the order of its plasma frequency. For metals, for instance, the typical value of this frequency is $\omega_{\rm pl} \approx 10^{16}\,{\rm s}^{-1}$~\cite{Ashcroft-book}, which is well below the frequency corresponding to the mirrors' rest energy, for typical values of the mass of mechanical components in state-of-the-art optomechanical devices~\cite{Aspelmeyer_RMP}. This means that the energies for which the model is pathologic lie well outside the domain of physical interest. Therefore, the theory can be regularized by introducing a frequency cut-off in Eq.~\eqref{Fn} of the order of the plasma frequency of the materials~\cite{Giulio-PRL-2013,Montalbano-2023}.

\subsection{Field ladder operators formalism\label{Sec:ladder}}
We have derived the quadratic expansion of the radiation pressure interaction in terms of the field  amplitudes $\hat{Q}_k$ and conjugate momenta $\hat{P}_k$. In this section, we recast the optomechanical Hamiltonian into a more familiar form by introducing the ladder operators $\hat{a}^\dag_{k0}$ and $\hat{a}_{k0}$. These are defined as creating and annihilating bare photons, respectively, that are excitations in the field modes corresponding to the equilibrium length $d$ of the cavity. As such, they take the form:
\begin{align}
	\hat{a}_{k0} &= \left(2\hbar\omega_{k0}\right)^{-1/2}\left(\omega_{k0} \hat{Q}_k + i \hat{P}_k\right),\label{a}\\
	\hat{a}_{k0}^\dag &= \left(2\hbar\omega_{k0}\right)^{-1/2}\left(\omega_{k0} \hat{Q}_k - i \hat{P}_k\right),\label{a_dag}
\end{align}
and satisfy the bosonic commutation relations $[a_{k0},a_{j0}^\dag] = \delta_{kj}$. By using the inverses of Eqs.~\eqref{a} and \eqref{a_dag}, the radiation pressure operators can be written as:
\begin{equation}
	\hat{F}_n = \frac{\hbar}{2}\sum_{kj} (-1)^{k+j}\sqrt{\omega_{k0} \omega_{j0}}\big(\hat{a}_k+\hat{a}^\dag_k\big)\big(\hat{a}_j+\hat{a}^\dag_j\big)^{(n)},\label{Fn_a}
\end{equation}
with $n=0,1$ within the quadratic approximation here considered. In Eq.~\eqref{Fn_a}, we defined the operators:
\begin{align}
	\hat{a}_k^{(0)} &\equiv \hat{a}_k, \label{a0}\\
	\hat{a}_k^{(n+1)} &\equiv \sum_j \sqrt{\frac{k}{j}} g_{jk} \hat{a}_j^{(n)} \qquad (n \in \mathbb{N}),\label{an}
\end{align}
which encode the same information as in Eqs.~\eqref{Q0} and \eqref{Qn}, but now expressed in the ladder operators formalism. 

It is now convenient to isolate the static effects generated by the field vacuum energy, thereby identifying in the rotated Hamiltonian $\hat{H}'$ the terms that account for dynamical effects of the radiation pressure. To this end, we decompose the radiation pressure operators $\hat{F}_n$ as sum of their normal ordered and vacuum energy contributions. By using the bosonic commutation relations introduced above, and indicating with $\normord{\hat{O}}$ the normal ordering of the generic field operator $\hat{O}$, these take the form:
\begin{align}
	\hat{F}_{0} &=\, \normord{\hat{F}_{0}} + \sum_k \frac{\hbar\omega_{k0}}{2},\label{F0_no}\\
	\hat{F}_{1} &=\, \normord{\hat{F}_{1}} + \frac{\hbar}{2}\sum_{kj}{}^{'} \frac{\omega_{k0}\omega_{j0}}{\omega_{k0}+\omega_{j0}},\label{F1_no}
\end{align}
where the primed sum indicates that the summation is over values of the indices such that $k\neq j$. By using Eqs.~\eqref{F0_no} and \eqref{F1_no}, the interaction Hamiltonian $\hat{H}_{\rm int}$ defined in Eq.~\eqref{Hint} can be decomposed in its vacuum $\hat{H}_{\rm int}^{\rm vac}$ and normal ordered $\hat{H}_{\rm int}^{\rm no}$ contributes:
\begin{equation}
	\hat{H}_{\rm int} = \hat{H}_{\rm int}^{\rm vac} + \hat{H}_{\rm int}^{\rm no},
\end{equation}
with
\begin{multline}
		\hat{H}_{\rm int}^{\rm vac} = \bigg(-\frac{\hat{x}}{d}+\frac{\hat{x}^2}{d^2}\bigg)\sum_k \frac{\hbar\omega_{k0}}{2} \\
		+ \frac{\hat{x}^2}{2d^2} \sum_{kj} \frac{\omega_{k0} \omega_{j0}}{\omega_{k0} + \omega_{j0}}, \label{Hint_vac}
\end{multline}
and	
\begin{equation}
		\hat{H}_{\rm int}^{\rm no} = -\frac{\hat{x}}{d} \normord{\hat{F}_0} + \frac{x^2}{d^2}\bigg( \frac{3}{2} \normord{\hat{F}_0} + \normord{\hat{F}_1}\bigg). \label{Hint_no}
\end{equation}
Combining the vacuum energy from the bare field Hamiltonian $\hat{H}_f = \sum_k \hbar\omega_k (\hat{a}_k^\dag \hat{a}_k + 1/2)$, with the first term in Eq.~\eqref{Hint_vac}, we obtain the quadratic expansion of the bare vacuum energy of the field in the configuration with the fluctuation boundary condition, that is:
\begin{equation}
	\sum_k \frac{\hbar\omega_{k0}}{2} \bigg(1-\frac{\hat{x}}{d}+\frac{\hat{x}^2}{d^2}\bigg) \approx \sum_k \frac{\hbar\omega_k(q)}{2}.
\end{equation}
This (static) divergent term, which is at the origin of the Casimir force, can be renormalized by subtracting the vacuum energy in the free space limit $q\to \infty$. This procedure leads to the well-known one-dimensional Casimir potential $-\hbar\pi/(24 q)$~\cite{Law1995}. The second vacuum term in Eq.~\eqref{Hint_vac} provides instead a static shift to the bare mechanical oscillation frequency. Such a divergent term can be absorbed in the definition of a renormalized frequency of the mechanical vibrations, which defines the oscillation frequency experimentally measured:
\begin{equation}
	\Omega_{\rm ren}^2 = \Omega^2 + \frac{1}{md^2}\sum_{kj} \frac{\omega_{k0} \omega_{j0}}{\omega_{k0} + \omega_{j0}}.
\end{equation}

After renormalizing the static vacuum energy contributes, the optomechanical Hamiltonian is finally given by:
\begin{equation}
	\hat{H}' = \frac{\hat{p}^2}{2m} + U(\hat{q}) + \sum_k \hbar\omega_{k0} \hat{a}_k^\dag \hat{a}_k  + \hat{H}_{\rm int}^{\rm no},
\end{equation}
where $U(\hat{q}) \equiv V_{\rm ren}(q)-\hbar\pi/(24q)$ is the renormalized confining potential which includes the Casimir energy. The dynamical effects of the radiation pressure interaction are therefore encoded in the standard (linear in $\hat{x}$) radiation force $f\equiv \normord{\hat{F}_0}/d$ originally derived by Law in \cite{Law1995}:
\begin{multline}
 f = \frac{\hbar}{2d}\sum_{kj} (-1)^{k+j}\sqrt{\omega_{k0} \omega_{j0}}\\ \times \big(\hat{a}_k \hat{a}_j+\hat{a}^\dag_k \hat{a}_j +\hat{a}^\dag_j \hat{a}_k +\hat{a}^\dag_j \hat{a}_k^\dag \big),\label{f}
\end{multline}
and in the dynamical frequency shift $\Delta\Omega^2 \equiv 2\normord{\hat{F}_1}/(m d^2)$, having the explicit form:
\begin{multline}
 \Delta\hat{\Omega}^2 = \frac{\hbar}{md^2}\sum_{kj} (-1)^{k+j}\sqrt{\omega_{k0} \omega_{j0}}\\ \times \bigg[\sum_s g_{sj}\sqrt{\frac{j}{s}}\big(\hat{a}_k \hat{a}_s +\hat{a}^\dag_k \hat{a}_s +\hat{a}^\dag_s \hat{a}_k +\hat{a}^\dag_k \hat{a}_s^\dag \big)\bigg].\label{Delta}
\end{multline}
This term, which accounts for the effects of quadratic fluctuations of the moving mirror around its equilibrium position, is the new result of this work (Note that we disregarded the effects of the quadratic term proportional to $\normord{\hat{F}_0}$ since, in the next section, we demonstrate these being negligible). By neglecting the Casimir potential, the complete mirror-field Hamiltonian, up to quadratic order in the mirror's fluctuations, takes therefore the form:
\begin{equation}
 H' = \frac{\hat{p}^2}{2m} + \frac{1}{2} m \big( \Omega_{\rm ren}^2 + \Delta\hat{\Omega}^2 \big)\hat{x}^2 + \sum_k \hbar\omega_k \hat{a}_k^\dag \hat{a}_k -\hat{x}\hat{f}.\label{H'fin}
\end{equation}
Eqs.~\eqref{f}-\eqref{H'fin} define the complete optomechanical Hamiltonian expanded up to quadratic order in the mechanical fluctuations.

\section{Discussion and conclusions\label{Sec:Conclusions}}
We finally examine in more details the quadratic corrections to the optomechanical Hamiltonian derived in the previous sections, focusing on their relevance in the context of current experimental setups. To this end, we perform an order of magnitude analysis of each of the interaction terms in Eq.~\eqref{Hint}, taking the quantum of mechanical energy $\hbar\Omega$ as a reference scale (by $\Omega$ it is here intended the physical, that is renormalized, mechanical oscillation frequency). 
By using $\omega_{\rm pl}$ as cut-off frequency for the sums in Eq.~\eqref{Fn_a}, we obtain the following ratio between the standard radiation pressure interaction strength and the mechanical energy:
\begin{equation}
	\frac{({\hat{x}}/{d})\hat{F}_0}{\hbar\Omega} \sim \left(\frac{x_{\text{zpf}}}{d}\right)\left(\frac{\omega_{\rm pl}}{\Omega}\right)  \equiv \lambda.\label{Ord1}
\end{equation}
Here, $x_{\rm zpf} = [{\hbar/(2m\Omega)}]^{1/2}$ is the amplitude of the zero-point (vacuum) fluctuations of the mirror's position within the confining harmonic potential. By definition, the parameter $\lambda$ accounts for the relative strength between the single-photon radiation pressure interaction and the unperturbed mechanical energy. Except that in the so-called \emph{ultra-strong coupling} regime~\cite{UltraStrong_Review}, in which $\lambda \approx 1$, in both the \emph{weak coupling} and \emph{strong coupling} limits it is verified that $\lambda \ll 1$. A perturbative treatment of the radiation pressure interaction is therefore permitted in these latter cases.

Given the result in Eq.~\eqref{Ord1}, it is straightforward to infer that the quadratic correction term involving the standard radiation pressure operator is of order:
\begin{equation}
	\frac{(\hat{x}^2/d^2)\hat{F}_0}{\hbar\Omega} \sim \left(\frac{x_{\text{zpf}}}{d}\right) \lambda.\label{Ord2a}
\end{equation}
For in state-of-the-art optomehcanical and electromechanical devices it is verified that $x_{\rm zpf}/d\ll 1$, this term can be safely neglected in the Hamiltonian.

What remains to be analized is the correction term involving the first-order contribution to the radiation pressure, that is the term: $(x^2/d^2) \hat{F}_1$. As described in the previous sections, $\hat{F}_1$ accounts for a higher-order effect of the radiation pressure and is therefore affected by a stronger ultraviolet divergence upon integration over the field's modes, compared to the standard radiation pressure $F_0$. Upon careful inspection, and using again $\omega_{\rm pl}$ as cut-off frequency, we deduce that:
\begin{equation}
	\frac{({\hat{x}}^2/{d}^2)\hat{F}_1}{\hbar\Omega} \sim \left(\frac{\Omega}{\omega_{10}}\right)\lambda^2,\label{Ord2b}
\end{equation}
where $\omega_{10}$ is the fundamental frequency of the cavity. This result suggests that, when working with high-frequency mechanical oscillators (or their circuital analogs) for which $\Omega\approx\omega_{10}$, this interaction term is of order $\lambda^2$.
Therefore, its effects need to be taken into account when investigating phenomena that are quadratic or higher in the mirror-field interaction. For example, this term is expected to modify the optomechanical vacuum, providing a non-zero contribution to the dressing process~\cite{Giulio-PRL-2013,Armata-PRD-2015}. Dynamical scenarios where nonlinear effects are significant have also been explored, where the quadratic correction to the optomechanical Hamiltonian likely provide a non-trivial contribution. These include optomechanical back-reaction phenomena~\cite{Savasta-PRX-2018,Butera-PRA-2019,Butera-EPL-2019,Ferreri-2024}, fluctuation mediated transfer of mechanical energy via the electromagnetic vacuum~\cite{Savasta-PRL-2019,Butera-InfFunc-2022}, deterministic generation of multi-photon entanglement by mechanical vibrations~\cite{Savasta-Hopping-2023,Savasta-SciPost-2025}, and quantum sensing~\cite{Schneiter-PRA-2020}.

We conclude by noticing that the operator $\hat{F}_1$ is identically zero when considering single field mode configurations. However, such an approximation break down for high-frequency mechanical resonators, where the oscillation frequency is comparable to the frequency spacing of the cavity modes. In this regime the contribution of $\hat{F}_1$ is expected to be significant and should be taken into account.


\section{Acknowledgments}
Enlightening discussions with Iacopo Carusotto, Stephen Barnett and the Quantum Theory Group at the University of Glasgow are warmly acknowledged. This research was supported by the University of Glasgow via the Lord Kelvin/Adam Smith (LKAS) Leadership Fellowship.

\appendix

\section{Exact mirror-field Hamiltonian in the ladder operators formalism\label{app:A}}
For completeness, we report here the exact Hamiltonian that describes the mirror-field interaction, written in terms of the position-dependent annihilation $\hat{a}_k(\hat{q})$ and creation $\hat{a}_k^\dag(\hat{q})$ operators for the field modes. These are defined as:
\begin{align}
	a_{k}(\hat{q}) &= \left[2\hbar\omega_{k}(\hat{q})\right]^{-1/2}\left[\omega_{k}(\hat{q}) Q_k + i P_k\right],\label{aq}\\
	a_{k}^\dag(\hat{q}) &= \left[2\hbar\omega_{k}(\hat{q})\right]^{-1/2}\left[\omega_{k}(\hat{q}) Q_k - i P_k\right],\label{aq_dag}
\end{align}
where $\omega_k(\hat{q})$ are the mirror-position-dependent eigenfrequencies of the field, defined in the main text. By using the inverse of Eqs.~\eqref{aq} and \eqref{aq_dag}, the Hamiltonian in Eq.~\eqref{Hq}, along with the non-adiabatic mirror-field coupling operator $\hat{\Gamma}(\hat{q})$, are written as~\cite{Law1995}:
\begin{multline}
	\hat{H} = \frac{\left[\hat{p}+\Gamma(\hat{q})\right]^2}{2m} + V(\hat{q}) - \frac{\hbar\pi}{24q} \\
	+\hbar \sum_k \omega(\hat{q})\hat{a}_k^\dag(\hat{q})\hat{a}_k(\hat{q}),\label{Ha}
\end{multline}
and
\begin{multline}
	\Gamma(\hat{q}) = \frac{i\hbar}{2q}\sum_{kj} g_{kj}\sqrt{\frac{k}{j}}\Big[\hat{a}_k^\dag(\hat{q})\hat{a}_j^\dag(\hat{q}) - \hat{a}_k(\hat{q})\hat{a}_j(\hat{q}) \\ + \hat{a}_k^\dag(\hat{q})\hat{a}_j(\hat{q})  - \hat{a}_j^\dag(\hat{q})\hat{a}_k(\hat{q})\Big].\label{Ga}
\end{multline}
Notice that, in Eq.~\eqref{Ha}, the vacuum field energy has been renormalized, resulting in the Casimir potential. Following Law's original approach~\cite{Law1995}, the Eq.~\eqref{H'fin} in the main text, that is the optomechanical Hamiltonian expanded up to quadratic order in the mirror's fluctuations, can be equivalently obtained by substituting in Eq.~\eqref{Ha} and \eqref{Ga} the following expansions for the annihilation and creation operators:
\begin{align}
	\hat{a}_k({q}) &\approx \hat{a}_{k0} - \frac{x}{2} \hat{a}_{k0}^\dag + \frac{\hat{x}^2}{4}\bigg(\hat{a}_{k0}^\dag + \frac{\hat{a}_{k0}}{2}\bigg),\\
	\hat{a}_k^\dag({q}) &\approx  \hat{a}_{k0}^\dag - \frac{x}{2} \hat{a}_{k0} + \frac{\hat{x}^2}{4}\bigg(\hat{a}_{k0} + \frac{\hat{a}_{k0}^\dag}{2}\bigg),
\end{align}
and performing the gauge transformation in Eq.~\eqref{T}.

\begin{widetext}

\section{Derivation of the gauge-transformed Hamiltonian\label{app:B}}
For the convenience of the reader, we outline the main steps in deriving of the gauge-transformed Hamiltonian reported in Eqs.~\eqref{H'}-\eqref{Hint} of the main text. For clarity, we first rewrite the expanded Hamiltonian introduced in Eq.~\eqref{Eq:Law_H_2nd} before transformation:
\begin{equation}
\hat{H} = \frac{\hat{p}_{\rm kin}^2}{2m} + V(\hat{x}) + \frac{1}{2}\sum_k \big(\hat{P}_k^2+\omega_{k0}^2 \hat{Q}_k^2\big) \\- \frac{\hat{x}}{d} \sum_k\omega_{k0}^2 \hat{Q}_k^2 + \frac{3\hat{x}^2}{2d^2} \sum_k\omega_{k0}^2 \hat{Q}_k^2. \label{A:Law_H_2nd}
\end{equation}
In Eq.~\eqref{A:Law_H_2nd}, we remind that the kinetic momentum operator of the mirror is defined as $\hat{p}_{\rm kin}=\hat{p}+\hat{\Gamma}_0\left(1-\hat{x}/{d}\right)$, with $\hat{p}$ the corresponding canonical momentum. It is straightforward to show that these are related by the transformation $\hat{T}(\hat{x}) = \exp \big[-i ({\hat{x}}/{\hbar})\big(1-{\hat{x}}/{2d}\big)\hat{\Gamma}_0\big]$, as:
\begin{equation}
\begin{split}
	\hat{T}^\dag(\hat{x}) \hat{p}_{\rm kin} \hat{T}(\hat{x}) &= \hat{T}^\dag(\hat{x}) \bigg[\hat{p}+\hat{\Gamma}_0\left(1-\frac{\hat{x}}{d}\right)\bigg] \hat{T}(\hat{x}) = \hat{p}.
\end{split}
\end{equation}
This result follows readily by working with the position representation of operators, in which $\hat{p} \equiv -i\hbar({d}/{dx})$. The potential $V(\hat{x})$ remains unchanged under $\hat{T}(\hat{x})$, since it depends solely on the mirror's position and thus commutes with the transformation itself.

Calculating the transformed field contributions in Eq.~\eqref{A:Law_H_2nd} is more involved. To this end, let us collectively indicate
\begin{equation}
	\hat{G} (\hat{x}) \equiv -\frac{\hat{x}}{\hbar}\left(1-\frac{\hat{x}}{2d}\right)\hat{\Gamma}_0 = \frac{\hat{x}}{\hbar d} \left(1-\frac{\hat{x}}{2d}\right) \bigg(\sum_{k}\hat{P}_k \hat{Q}_k^{(1)}\bigg),
\end{equation}
so that the gauge transformation can be written in the compact form $\hat{T}(\hat{x}) = \exp\big[i \hat{G}(\hat{x})\big]$. Working to quadratic order in $\hat{x}$, the transformation of the field mode amplitudes is given by
\begin{equation}
\hat{T}^\dag(\hat{x})\hat{Q}_k \hat{T}(\hat{x}) = e^{ -i \hat{G}(\hat{x})} \hat{Q}_k e^{ i \hat{G}(\hat{x})} \approx\bigg(1-i \hat{G} +\frac{i^2}{2} \hat{G}^2\bigg) \hat{Q}_k \bigg(1+i \hat{G} +\frac{i^2}{2} \hat{G}^2\bigg),
\end{equation}
that, upon evaluating the commutators, yields:
\begin{equation}
	\hat{T}^\dag(\hat{x})\hat{Q}_k \hat{T}(\hat{x}) = \hat{Q}_k - \frac{\hat{x}}{d} \left(1-\frac{\hat{x}}{2d}\right)\hat{Q}_k^{(1)} + \frac{\hat{x}^2}{2d}\hat{Q}_k^{(2)}.\label{Qs_transf}
\end{equation}
Similarly, the conjugate momenta transform as:
\begin{equation}
	\hat{T}^\dag(\hat{x})\hat{P}_k \hat{T}(\hat{x}) = \hat{P}_k - \frac{\hat{x}}{d} \left(1-\frac{\hat{x}}{2d}\right)\hat{P}_k^{(1)} + \frac{\hat{x}^2}{2d}\hat{P}_k^{(2)}.\label{Ps_transf}
\end{equation}

Eqs.~\eqref{Qs_transf} and \eqref{Ps_transf} can now be employed to transform the field's contributes in Eq.~\eqref{A:Law_H_2nd}. 
Specifically, for the field kinetic energy:
\begin{equation}
	\hat{T}^\dag(\hat{x}) \bigg(\frac{1}{2}\sum_k \hat{P}_k^2\bigg) \hat{T}(\hat{x}) = \frac{1}{2}\sum_k \hat{P}_k^2, \label{A:1}
\end{equation}
for the field potential energy:
\begin{equation}
	\hat{T}^\dag(\hat{x}) \bigg(\frac{1}{2}\sum_k \omega_{k0}^2 \hat{Q}_k^2\bigg) \hat{T}(\hat{x}) = \frac{1}{2}\sum_k \omega_{k0}^2 \hat{Q}_k^2 -\frac{\hat{x}}{d}\bigg(1-\frac{\hat{x}}{2d}\bigg)\sum_k\omega_{k0}\hat{Q}_k \hat{Q}_k^{(1)} + \frac{\hat{x}^2}{2d^2}\sum_k^2 \bigg(\hat{Q}_k \hat{Q}_k^{(2)} + [\hat{Q}_k^{(1)}]^2\bigg),\label{A:2}
\end{equation}
for the term linear in $\hat{x}$:
\begin{equation}
	\hat{T}^\dag(\hat{x})\bigg(-\frac{\hat{x}}{d}\sum_k\omega_{k0}^2 \hat{Q}_k^2\bigg) \hat{T}(\hat{x}) = -\frac{\hat{x}}{d} \sum_k\omega_{k0}^2 \hat{Q}_k^2 + 2\frac{\hat{x}^2}{d^2} \sum_k\omega_{k0}^2 \hat{Q}_k \hat{Q}_k^{(1)},\label{A:3}
\end{equation}
and, finally, for the term quadratic in $\hat{x}$:
\begin{equation}
\hat{T}^\dag(\hat{x})\bigg( \frac{3}{2}\frac{\hat{x}^2}{d^2} \sum_s\omega_{k0}^2 \hat{Q}_k^2 \bigg) \hat{T}(\hat{x}) = \frac{3}{2}\frac{\hat{x}^2}{d^2} \sum_k\omega_{k0}^2 \hat{Q}_k^2.\label{A:4}
\end{equation}
Combining these results, we obtain after some algebra:
\begin{equation}
	\frac{1}{2} \sum_k \left(\hat{P}_{k0}^2 + \omega_{k0}^2 \hat{Q}_k^2\right) -\frac{\hat{x}}{d} \hat{F}_0 + \frac{\hat{x}^2}{d^2}\left(\frac{3}{2}\hat{F}_0 + \hat{F}_1\right),
\end{equation}
which, together with the transformed mirror's Hamiltonian derived before, produces the final gauge-transformed Hamiltonian given in Eqs.~\eqref{H'}-\eqref{Hint}.
\end{widetext}


\bibliography{TwoMirrors.bib}

\begin{thebibliography}{81}%
\makeatletter
\providecommand \@ifxundefined [1]{%
 \@ifx{#1\undefined}
}%
\providecommand \@ifnum [1]{%
 \ifnum #1\expandafter \@firstoftwo
 \else \expandafter \@secondoftwo
 \fi
}%
\providecommand \@ifx [1]{%
 \ifx #1\expandafter \@firstoftwo
 \else \expandafter \@secondoftwo
 \fi
}%
\providecommand \natexlab [1]{#1}%
\providecommand \enquote  [1]{``#1''}%
\providecommand \bibnamefont  [1]{#1}%
\providecommand \bibfnamefont [1]{#1}%
\providecommand \citenamefont [1]{#1}%
\providecommand \href@noop [0]{\@secondoftwo}%
\providecommand \href [0]{\begingroup \@sanitize@url \@href}%
\providecommand \@href[1]{\@@startlink{#1}\@@href}%
\providecommand \@@href[1]{\endgroup#1\@@endlink}%
\providecommand \@sanitize@url [0]{\catcode `\\12\catcode `\$12\catcode
  `\&12\catcode `\#12\catcode `\^12\catcode `\_12\catcode `\%12\relax}%
\providecommand \@@startlink[1]{}%
\providecommand \@@endlink[0]{}%
\providecommand \url  [0]{\begingroup\@sanitize@url \@url }%
\providecommand \@url [1]{\endgroup\@href {#1}{\urlprefix }}%
\providecommand \urlprefix  [0]{URL }%
\providecommand \Eprint [0]{\href }%
\providecommand \doibase [0]{https://doi.org/}%
\providecommand \selectlanguage [0]{\@gobble}%
\providecommand \bibinfo  [0]{\@secondoftwo}%
\providecommand \bibfield  [0]{\@secondoftwo}%
\providecommand \translation [1]{[#1]}%
\providecommand \BibitemOpen [0]{}%
\providecommand \bibitemStop [0]{}%
\providecommand \bibitemNoStop [0]{.\EOS\space}%
\providecommand \EOS [0]{\spacefactor3000\relax}%
\providecommand \BibitemShut  [1]{\csname bibitem#1\endcsname}%
\let\auto@bib@innerbib\@empty
\bibitem [{\citenamefont {Aspelmeyer}\ \emph {et~al.}(2014)\citenamefont
  {Aspelmeyer}, \citenamefont {Kippenberg},\ and\ \citenamefont
  {Marquardt}}]{Aspelmeyer_RMP}%
  \BibitemOpen
  \bibfield  {author} {\bibinfo {author} {\bibfnamefont {M.}~\bibnamefont
  {Aspelmeyer}}, \bibinfo {author} {\bibfnamefont {T.~J.}\ \bibnamefont
  {Kippenberg}},\ and\ \bibinfo {author} {\bibfnamefont {F.}~\bibnamefont
  {Marquardt}},\ }\bibfield  {title} {\bibinfo {title} {Cavity optomechanics},\
  }\href {https://doi.org/10.1103/RevModPhys.86.1391} {\bibfield  {journal}
  {\bibinfo  {journal} {Rev. Mod. Phys.}\ }\textbf {\bibinfo {volume} {86}},\
  \bibinfo {pages} {1391} (\bibinfo {year} {2014})}\BibitemShut {NoStop}%
\bibitem [{\citenamefont {Biermann}(1951)}]{Comet1951}%
  \BibitemOpen
  \bibfield  {author} {\bibinfo {author} {\bibfnamefont {L.}~\bibnamefont
  {Biermann}},\ }\bibfield  {title} {\bibinfo {title} {Kometenschweife und
  solare korpuskularstrahlung},\ }\href@noop {} {\bibfield  {journal} {\bibinfo
   {journal} {Z Astrophys}\ }\textbf {\bibinfo {volume} {29}},\ \bibinfo
  {pages} {274} (\bibinfo {year} {1951})}\BibitemShut {NoStop}%
\bibitem [{\citenamefont {Ashkin}(1978)}]{Ashkin1978}%
  \BibitemOpen
  \bibfield  {author} {\bibinfo {author} {\bibfnamefont {A.}~\bibnamefont
  {Ashkin}},\ }\bibfield  {title} {\bibinfo {title} {Trapping of atoms by
  resonance radiation pressure},\ }\href
  {https://doi.org/10.1103/PhysRevLett.40.729} {\bibfield  {journal} {\bibinfo
  {journal} {Phys. Rev. Lett.}\ }\textbf {\bibinfo {volume} {40}},\ \bibinfo
  {pages} {729} (\bibinfo {year} {1978})}\BibitemShut {NoStop}%
\bibitem [{\citenamefont {Ashkin}\ \emph {et~al.}(1986)\citenamefont {Ashkin},
  \citenamefont {Dziedzic}, \citenamefont {Bjorkholm},\ and\ \citenamefont
  {Chu}}]{Ashkin1986}%
  \BibitemOpen
  \bibfield  {author} {\bibinfo {author} {\bibfnamefont {A.}~\bibnamefont
  {Ashkin}}, \bibinfo {author} {\bibfnamefont {J.~M.}\ \bibnamefont
  {Dziedzic}}, \bibinfo {author} {\bibfnamefont {J.~E.}\ \bibnamefont
  {Bjorkholm}},\ and\ \bibinfo {author} {\bibfnamefont {S.}~\bibnamefont
  {Chu}},\ }\bibfield  {title} {\bibinfo {title} {Observation of a single-beam
  gradient force optical trap for dielectric particles},\ }\href@noop {}
  {\bibfield  {journal} {\bibinfo  {journal} {Optics letters}\ }\textbf
  {\bibinfo {volume} {11}},\ \bibinfo {pages} {288} (\bibinfo {year}
  {1986})}\BibitemShut {NoStop}%
\bibitem [{\citenamefont {Ashkin}\ and\ \citenamefont
  {Dziedzic}(1987)}]{Ashkin1987}%
  \BibitemOpen
  \bibfield  {author} {\bibinfo {author} {\bibfnamefont {A.}~\bibnamefont
  {Ashkin}}\ and\ \bibinfo {author} {\bibfnamefont {J.~M.}\ \bibnamefont
  {Dziedzic}},\ }\bibfield  {title} {\bibinfo {title} {Optical trapping and
  manipulation of viruses and bacteria},\ }\href@noop {} {\bibfield  {journal}
  {\bibinfo  {journal} {Science}\ }\textbf {\bibinfo {volume} {235}},\ \bibinfo
  {pages} {1517} (\bibinfo {year} {1987})}\BibitemShut {NoStop}%
\bibitem [{\citenamefont {Chu}\ and\ \citenamefont
  {Wieman}(1989)}]{LaserCooling}%
  \BibitemOpen
  \bibfield  {author} {\bibinfo {author} {\bibfnamefont {S.}~\bibnamefont
  {Chu}}\ and\ \bibinfo {author} {\bibfnamefont {C.}~\bibnamefont {Wieman}},\
  }\bibfield  {title} {\bibinfo {title} {Laser cooling and trapping of atoms:
  Introduction},\ }\href {https://doi.org/10.1364/JOSAB.6.002020} {\bibfield
  {journal} {\bibinfo  {journal} {J. Opt. Soc. Am. B}\ }\textbf {\bibinfo
  {volume} {6}},\ \bibinfo {pages} {2020} (\bibinfo {year} {1989})}\BibitemShut
  {NoStop}%
\bibitem [{\citenamefont {Chan}\ \emph {et~al.}(2011)\citenamefont {Chan},
  \citenamefont {Alegre}, \citenamefont {Safavi-Naeini}, \citenamefont {Hill},
  \citenamefont {Krause}, \citenamefont {Gr{\"o}blacher}, \citenamefont
  {Aspelmeyer},\ and\ \citenamefont {Painter}}]{chan2011laser}%
  \BibitemOpen
  \bibfield  {author} {\bibinfo {author} {\bibfnamefont {J.}~\bibnamefont
  {Chan}}, \bibinfo {author} {\bibfnamefont {T.}~\bibnamefont {Alegre}},
  \bibinfo {author} {\bibfnamefont {A.}~\bibnamefont {Safavi-Naeini}}, \bibinfo
  {author} {\bibfnamefont {J.}~\bibnamefont {Hill}}, \bibinfo {author}
  {\bibfnamefont {A.}~\bibnamefont {Krause}}, \bibinfo {author} {\bibfnamefont
  {S.}~\bibnamefont {Gr{\"o}blacher}}, \bibinfo {author} {\bibfnamefont
  {M.}~\bibnamefont {Aspelmeyer}},\ and\ \bibinfo {author} {\bibfnamefont
  {O.}~\bibnamefont {Painter}},\ }\bibfield  {title} {\bibinfo {title} {Laser
  cooling of a nanomechanical oscillator into its quantum ground state},\
  }\href {https://doi.org/10.1038/nature10461} {\bibfield  {journal} {\bibinfo
  {journal} {Nature}\ }\textbf {\bibinfo {volume} {478}},\ \bibinfo {pages}
  {89} (\bibinfo {year} {2011})}\BibitemShut {NoStop}%
\bibitem [{\citenamefont {Deli{\`c}}\ \emph {et~al.}(2020)\citenamefont
  {Deli{\`c}}, \citenamefont {Reisenbauer}, \citenamefont {Dare}, \citenamefont
  {Grass}, \citenamefont {Vuleti\`{c}}, \citenamefont {Kiesel},\ and\
  \citenamefont {Aspelmeyer}}]{Aspelmeyer-Science-2020}%
  \BibitemOpen
  \bibfield  {author} {\bibinfo {author} {\bibfnamefont {U.}~\bibnamefont
  {Deli{\`c}}}, \bibinfo {author} {\bibfnamefont {M.}~\bibnamefont
  {Reisenbauer}}, \bibinfo {author} {\bibfnamefont {K.}~\bibnamefont {Dare}},
  \bibinfo {author} {\bibfnamefont {D.}~\bibnamefont {Grass}}, \bibinfo
  {author} {\bibfnamefont {V.}~\bibnamefont {Vuleti\`{c}}}, \bibinfo {author}
  {\bibfnamefont {N.}~\bibnamefont {Kiesel}},\ and\ \bibinfo {author}
  {\bibfnamefont {M.}~\bibnamefont {Aspelmeyer}},\ }\bibfield  {title}
  {\bibinfo {title} {Cooling of a levitated nanoparticle to the motional
  quantum ground state},\ }\href {https://doi.org/10.1126/science.aba3993}
  {\bibfield  {journal} {\bibinfo  {journal} {Science}\ }\textbf {\bibinfo
  {volume} {367}},\ \bibinfo {pages} {892} (\bibinfo {year}
  {2020})}\BibitemShut {NoStop}%
\bibitem [{\citenamefont {Abramovici}\ \emph {et~al.}(1992)\citenamefont
  {Abramovici}, \citenamefont {Althouse}, \citenamefont {Drever}, \citenamefont
  {G\"ursel}, \citenamefont {Kawamura}, \citenamefont {Raab}, \citenamefont
  {Shoemaker}, \citenamefont {Sievers}, \citenamefont {Spero}, \citenamefont
  {Thorne}, \citenamefont {Vogt}, \citenamefont {Weiss}, \citenamefont
  {Whitcomb},\ and\ \citenamefont {Zucker}}]{Abramovici-science-1992}%
  \BibitemOpen
  \bibfield  {author} {\bibinfo {author} {\bibfnamefont {A.}~\bibnamefont
  {Abramovici}}, \bibinfo {author} {\bibfnamefont {W.~E.}\ \bibnamefont
  {Althouse}}, \bibinfo {author} {\bibfnamefont {R.~W.~P.}\ \bibnamefont
  {Drever}}, \bibinfo {author} {\bibfnamefont {Y.}~\bibnamefont {G\"ursel}},
  \bibinfo {author} {\bibfnamefont {S.}~\bibnamefont {Kawamura}}, \bibinfo
  {author} {\bibfnamefont {F.~J.}\ \bibnamefont {Raab}}, \bibinfo {author}
  {\bibfnamefont {D.}~\bibnamefont {Shoemaker}}, \bibinfo {author}
  {\bibfnamefont {L.}~\bibnamefont {Sievers}}, \bibinfo {author} {\bibfnamefont
  {R.~E.}\ \bibnamefont {Spero}}, \bibinfo {author} {\bibfnamefont {K.~S.}\
  \bibnamefont {Thorne}}, \bibinfo {author} {\bibfnamefont {R.~E.}\
  \bibnamefont {Vogt}}, \bibinfo {author} {\bibfnamefont {R.}~\bibnamefont
  {Weiss}}, \bibinfo {author} {\bibfnamefont {S.~E.}\ \bibnamefont
  {Whitcomb}},\ and\ \bibinfo {author} {\bibfnamefont {M.~E.}\ \bibnamefont
  {Zucker}},\ }\bibfield  {title} {\bibinfo {title} {Ligo: The laser
  interferometer gravitational-wave observatory},\ }\href
  {https://doi.org/10.1126/science.256.5055.325} {\bibfield  {journal}
  {\bibinfo  {journal} {Science}\ }\textbf {\bibinfo {volume} {256}},\ \bibinfo
  {pages} {325} (\bibinfo {year} {1992})}\BibitemShut {NoStop}%
\bibitem [{\citenamefont {Accadia}\ \emph {et~al.}(2012)\citenamefont
  {Accadia}, \citenamefont {Acernese}, \citenamefont {Alshourbagy},
  \citenamefont {Amico}, \citenamefont {Antonucci}, \citenamefont {Aoudia},
  \citenamefont {Arnaud}, \citenamefont {Arnault}, \citenamefont {Arun},
  \citenamefont {Astone} \emph {et~al.}}]{virgo-2012}%
  \BibitemOpen
  \bibfield  {author} {\bibinfo {author} {\bibfnamefont {T.}~\bibnamefont
  {Accadia}}, \bibinfo {author} {\bibfnamefont {F.}~\bibnamefont {Acernese}},
  \bibinfo {author} {\bibfnamefont {M.}~\bibnamefont {Alshourbagy}}, \bibinfo
  {author} {\bibfnamefont {P.}~\bibnamefont {Amico}}, \bibinfo {author}
  {\bibfnamefont {F.}~\bibnamefont {Antonucci}}, \bibinfo {author}
  {\bibfnamefont {S.}~\bibnamefont {Aoudia}}, \bibinfo {author} {\bibfnamefont
  {N.}~\bibnamefont {Arnaud}}, \bibinfo {author} {\bibfnamefont
  {C.}~\bibnamefont {Arnault}}, \bibinfo {author} {\bibfnamefont
  {K.}~\bibnamefont {Arun}}, \bibinfo {author} {\bibfnamefont {P.}~\bibnamefont
  {Astone}}, \emph {et~al.},\ }\bibfield  {title} {\bibinfo {title} {Virgo: a
  laser interferometer to detect gravitational waves},\ }\href
  {https://doi.org/10.1088/1748-0221/7/03/P03012} {\bibfield  {journal}
  {\bibinfo  {journal} {Journal of instrumentation}\ }\textbf {\bibinfo
  {volume} {7}},\ \bibinfo {pages} {P03012} (\bibinfo {year}
  {2012})}\BibitemShut {NoStop}%
\bibitem [{\citenamefont {Caves}(1980)}]{Caves-PRL-1980}%
  \BibitemOpen
  \bibfield  {author} {\bibinfo {author} {\bibfnamefont {C.~M.}\ \bibnamefont
  {Caves}},\ }\bibfield  {title} {\bibinfo {title} {Quantum-mechanical
  radiation-pressure fluctuations in an interferometer},\ }\href
  {https://doi.org/10.1103/PhysRevLett.45.75} {\bibfield  {journal} {\bibinfo
  {journal} {Phys. Rev. Lett.}\ }\textbf {\bibinfo {volume} {45}},\ \bibinfo
  {pages} {75} (\bibinfo {year} {1980})}\BibitemShut {NoStop}%
\bibitem [{\citenamefont {Loudon}(1981)}]{Loudon-PRL-1981}%
  \BibitemOpen
  \bibfield  {author} {\bibinfo {author} {\bibfnamefont {R.}~\bibnamefont
  {Loudon}},\ }\bibfield  {title} {\bibinfo {title} {Quantum limit on the
  michelson interferometer used for gravitational-wave detection},\ }\href
  {https://doi.org/10.1103/PhysRevLett.47.815} {\bibfield  {journal} {\bibinfo
  {journal} {Phys. Rev. Lett.}\ }\textbf {\bibinfo {volume} {47}},\ \bibinfo
  {pages} {815} (\bibinfo {year} {1981})}\BibitemShut {NoStop}%
\bibitem [{\citenamefont {Meystre}\ \emph {et~al.}(1985)\citenamefont
  {Meystre}, \citenamefont {Wright}, \citenamefont {McCullen},\ and\
  \citenamefont {Vignes}}]{Vignes-1985}%
  \BibitemOpen
  \bibfield  {author} {\bibinfo {author} {\bibfnamefont {P.}~\bibnamefont
  {Meystre}}, \bibinfo {author} {\bibfnamefont {E.~M.}\ \bibnamefont {Wright}},
  \bibinfo {author} {\bibfnamefont {J.~D.}\ \bibnamefont {McCullen}},\ and\
  \bibinfo {author} {\bibfnamefont {E.}~\bibnamefont {Vignes}},\ }\bibfield
  {title} {\bibinfo {title} {Theory of radiation-pressure-driven
  interferometers},\ }\href {https://doi.org/10.1364/JOSAB.2.001830} {\bibfield
   {journal} {\bibinfo  {journal} {J. Opt. Soc. Am. B}\ }\textbf {\bibinfo
  {volume} {2}},\ \bibinfo {pages} {1830} (\bibinfo {year} {1985})}\BibitemShut
  {NoStop}%
\bibitem [{\citenamefont {Pace}\ \emph {et~al.}(1993)\citenamefont {Pace},
  \citenamefont {Collett},\ and\ \citenamefont {Walls}}]{Walls-PRA-1993}%
  \BibitemOpen
  \bibfield  {author} {\bibinfo {author} {\bibfnamefont {A.~F.}\ \bibnamefont
  {Pace}}, \bibinfo {author} {\bibfnamefont {M.~J.}\ \bibnamefont {Collett}},\
  and\ \bibinfo {author} {\bibfnamefont {D.~F.}\ \bibnamefont {Walls}},\
  }\bibfield  {title} {\bibinfo {title} {Quantum limits in interferometric
  detection of gravitational radiation},\ }\href
  {https://doi.org/10.1103/PhysRevA.47.3173} {\bibfield  {journal} {\bibinfo
  {journal} {Phys. Rev. A}\ }\textbf {\bibinfo {volume} {47}},\ \bibinfo
  {pages} {3173} (\bibinfo {year} {1993})}\BibitemShut {NoStop}%
\bibitem [{\citenamefont {O’Connell}\ \emph
  {et~al.}(2010{\natexlab{a}})\citenamefont {O’Connell}, \citenamefont
  {Hofheinz}, \citenamefont {Ansmann}, \citenamefont {Bialczak}, \citenamefont
  {Lenander}, \citenamefont {Lucero}, \citenamefont {Neeley}, \citenamefont
  {Sank}, \citenamefont {Wang}, \citenamefont {Weides}, \citenamefont {Wenner},
  \citenamefont {Martinis},\ and\ \citenamefont
  {Cleland}}]{OConnell-Science-2010}%
  \BibitemOpen
  \bibfield  {author} {\bibinfo {author} {\bibfnamefont {A.~D.}\ \bibnamefont
  {O’Connell}}, \bibinfo {author} {\bibfnamefont {M.}~\bibnamefont
  {Hofheinz}}, \bibinfo {author} {\bibfnamefont {M.}~\bibnamefont {Ansmann}},
  \bibinfo {author} {\bibfnamefont {R.}~\bibnamefont {Bialczak}}, \bibinfo
  {author} {\bibfnamefont {M.}~\bibnamefont {Lenander}}, \bibinfo {author}
  {\bibfnamefont {E.}~\bibnamefont {Lucero}}, \bibinfo {author} {\bibfnamefont
  {M.}~\bibnamefont {Neeley}}, \bibinfo {author} {\bibfnamefont
  {D.}~\bibnamefont {Sank}}, \bibinfo {author} {\bibfnamefont {H.}~\bibnamefont
  {Wang}}, \bibinfo {author} {\bibfnamefont {M.}~\bibnamefont {Weides}},
  \bibinfo {author} {\bibfnamefont {J.}~\bibnamefont {Wenner}}, \bibinfo
  {author} {\bibfnamefont {J.~M.}\ \bibnamefont {Martinis}},\ and\ \bibinfo
  {author} {\bibfnamefont {A.~N.}\ \bibnamefont {Cleland}},\ }\href
  {https://doi.org/10.1038/nature08967} {\bibfield  {journal} {\bibinfo
  {journal} {Nature}\ }\textbf {\bibinfo {volume} {464}},\ \bibinfo {pages}
  {697} (\bibinfo {year} {2010}{\natexlab{a}})}\BibitemShut {NoStop}%
\bibitem [{\citenamefont {Teufel}\ \emph
  {et~al.}(2011{\natexlab{a}})\citenamefont {Teufel}, \citenamefont {Donner},
  \citenamefont {Li}, \citenamefont {Harlow}, \citenamefont {Allman},
  \citenamefont {Cicak}, \citenamefont {Sirois}, \citenamefont {Whittaker},
  \citenamefont {Lehnert},\ and\ \citenamefont
  {Simmonds}}]{teufel2011sideband}%
  \BibitemOpen
  \bibfield  {author} {\bibinfo {author} {\bibfnamefont {J.~D.}\ \bibnamefont
  {Teufel}}, \bibinfo {author} {\bibfnamefont {T.}~\bibnamefont {Donner}},
  \bibinfo {author} {\bibfnamefont {D.}~\bibnamefont {Li}}, \bibinfo {author}
  {\bibfnamefont {J.~W.}\ \bibnamefont {Harlow}}, \bibinfo {author}
  {\bibfnamefont {M.~S.}\ \bibnamefont {Allman}}, \bibinfo {author}
  {\bibfnamefont {K.}~\bibnamefont {Cicak}}, \bibinfo {author} {\bibfnamefont
  {A.~J.}\ \bibnamefont {Sirois}}, \bibinfo {author} {\bibfnamefont {J.~D.}\
  \bibnamefont {Whittaker}}, \bibinfo {author} {\bibfnamefont {K.~W.}\
  \bibnamefont {Lehnert}},\ and\ \bibinfo {author} {\bibfnamefont {R.~W.}\
  \bibnamefont {Simmonds}},\ }\bibfield  {title} {\bibinfo {title} {Sideband
  cooling of micromechanical motion to the quantum ground state},\ }\href
  {https://doi.org/10.1038/nature10261} {\bibfield  {journal} {\bibinfo
  {journal} {Nature}\ }\textbf {\bibinfo {volume} {475}},\ \bibinfo {pages}
  {359} (\bibinfo {year} {2011}{\natexlab{a}})}\BibitemShut {NoStop}%
\bibitem [{\citenamefont {Safavi-Naeini}\ \emph {et~al.}(2012)\citenamefont
  {Safavi-Naeini}, \citenamefont {Chan}, \citenamefont {Hill}, \citenamefont
  {Alegre}, \citenamefont {Krause},\ and\ \citenamefont
  {Painter}}]{Painter2012}%
  \BibitemOpen
  \bibfield  {author} {\bibinfo {author} {\bibfnamefont {A.~H.}\ \bibnamefont
  {Safavi-Naeini}}, \bibinfo {author} {\bibfnamefont {J.}~\bibnamefont {Chan}},
  \bibinfo {author} {\bibfnamefont {J.~T.}\ \bibnamefont {Hill}}, \bibinfo
  {author} {\bibfnamefont {T.~P.~M.}\ \bibnamefont {Alegre}}, \bibinfo {author}
  {\bibfnamefont {A.}~\bibnamefont {Krause}},\ and\ \bibinfo {author}
  {\bibfnamefont {O.}~\bibnamefont {Painter}},\ }\bibfield  {title} {\bibinfo
  {title} {Observation of quantum motion of a nanomechanical resonator},\
  }\href {https://doi.org/10.1103/PhysRevLett.108.033602} {\bibfield  {journal}
  {\bibinfo  {journal} {Phys. Rev. Lett.}\ }\textbf {\bibinfo {volume} {108}},\
  \bibinfo {pages} {033602} (\bibinfo {year} {2012})}\BibitemShut {NoStop}%
\bibitem [{\citenamefont {Abbott}\ \emph {et~al.}(2009)\citenamefont {Abbott},
  \citenamefont {Abbott}, \citenamefont {Adhikari}, \citenamefont {Ajith},
  \citenamefont {Allen}, \citenamefont {Allen}, \citenamefont {Amin},
  \citenamefont {Anderson}, \citenamefont {Anderson}, \citenamefont {Arain}
  \emph {et~al.}}]{abbott2009observation}%
  \BibitemOpen
  \bibfield  {author} {\bibinfo {author} {\bibfnamefont {B.}~\bibnamefont
  {Abbott}}, \bibinfo {author} {\bibfnamefont {R.}~\bibnamefont {Abbott}},
  \bibinfo {author} {\bibfnamefont {R.}~\bibnamefont {Adhikari}}, \bibinfo
  {author} {\bibfnamefont {P.}~\bibnamefont {Ajith}}, \bibinfo {author}
  {\bibfnamefont {B.}~\bibnamefont {Allen}}, \bibinfo {author} {\bibfnamefont
  {G.}~\bibnamefont {Allen}}, \bibinfo {author} {\bibfnamefont
  {R.}~\bibnamefont {Amin}}, \bibinfo {author} {\bibfnamefont {S.~B.}\
  \bibnamefont {Anderson}}, \bibinfo {author} {\bibfnamefont {W.~G.}\
  \bibnamefont {Anderson}}, \bibinfo {author} {\bibfnamefont {M.~A.}\
  \bibnamefont {Arain}}, \emph {et~al.},\ }\bibfield  {title} {\bibinfo {title}
  {Observation of a kilogram-scale oscillator near its quantum ground state},\
  }\href {https://doi.org/10.1088/1367-2630/11/7/073032} {\bibfield  {journal}
  {\bibinfo  {journal} {New J. Phys.}\ }\textbf {\bibinfo {volume} {11}},\
  \bibinfo {pages} {073032} (\bibinfo {year} {2009})}\BibitemShut {NoStop}%
\bibitem [{\citenamefont {Barzanjeh}\ \emph {et~al.}(2022)\citenamefont
  {Barzanjeh}, \citenamefont {Xuereb}, \citenamefont {Gr{\"o}blacher},
  \citenamefont {Paternostro}, \citenamefont {Regal},\ and\ \citenamefont
  {Weig}}]{Paternostro_Review}%
  \BibitemOpen
  \bibfield  {author} {\bibinfo {author} {\bibfnamefont {S.}~\bibnamefont
  {Barzanjeh}}, \bibinfo {author} {\bibfnamefont {A.}~\bibnamefont {Xuereb}},
  \bibinfo {author} {\bibfnamefont {S.}~\bibnamefont {Gr{\"o}blacher}},
  \bibinfo {author} {\bibfnamefont {M.}~\bibnamefont {Paternostro}}, \bibinfo
  {author} {\bibfnamefont {C.}~\bibnamefont {Regal}},\ and\ \bibinfo {author}
  {\bibfnamefont {E.}~\bibnamefont {Weig}},\ }\bibfield  {title} {\bibinfo
  {title} {Optomechanics for quantum technologies},\ }\href
  {https://doi.org/10.1038/s41567-021-01402-0} {\bibfield  {journal} {\bibinfo
  {journal} {Nat. Phys.}\ }\textbf {\bibinfo {volume} {18}},\ \bibinfo {pages}
  {15} (\bibinfo {year} {2022})}\BibitemShut {NoStop}%
\bibitem [{\citenamefont {Stange}\ \emph {et~al.}(2021)\citenamefont {Stange},
  \citenamefont {Campbell},\ and\ \citenamefont
  {Bishop}}]{Stange-PhysToday-2021}%
  \BibitemOpen
  \bibfield  {author} {\bibinfo {author} {\bibfnamefont {A.}~\bibnamefont
  {Stange}}, \bibinfo {author} {\bibfnamefont {D.}~\bibnamefont {Campbell}},\
  and\ \bibinfo {author} {\bibfnamefont {D.}~\bibnamefont {Bishop}},\
  }\bibfield  {title} {\bibinfo {title} {Science and technology of the casimir
  effect},\ }\href {https://doi.org/10.1063/PT.3.4656} {\bibfield  {journal}
  {\bibinfo  {journal} {Phys. Today}\ }\textbf {\bibinfo {volume} {74}},\
  \bibinfo {pages} {42} (\bibinfo {year} {2021})}\BibitemShut {NoStop}%
\bibitem [{\citenamefont {Pistolesi}\ \emph {et~al.}(2021)\citenamefont
  {Pistolesi}, \citenamefont {Cleland},\ and\ \citenamefont
  {Bachtold}}]{MechQbit-PRX-2021}%
  \BibitemOpen
  \bibfield  {author} {\bibinfo {author} {\bibfnamefont {F.}~\bibnamefont
  {Pistolesi}}, \bibinfo {author} {\bibfnamefont {A.~N.}\ \bibnamefont
  {Cleland}},\ and\ \bibinfo {author} {\bibfnamefont {A.}~\bibnamefont
  {Bachtold}},\ }\bibfield  {title} {\bibinfo {title} {Proposal for a
  nanomechanical qubit},\ }\href {https://doi.org/10.1103/PhysRevX.11.031027}
  {\bibfield  {journal} {\bibinfo  {journal} {Phys. Rev. X}\ }\textbf {\bibinfo
  {volume} {11}},\ \bibinfo {pages} {031027} (\bibinfo {year}
  {2021})}\BibitemShut {NoStop}%
\bibitem [{\citenamefont {Navarathna}\ and\ \citenamefont
  {Bowen}(2022)}]{navarathna2022good}%
  \BibitemOpen
  \bibfield  {author} {\bibinfo {author} {\bibfnamefont {A.}~\bibnamefont
  {Navarathna}}\ and\ \bibinfo {author} {\bibfnamefont {W.~P.}\ \bibnamefont
  {Bowen}},\ }\bibfield  {title} {\bibinfo {title} {Good vibrations for quantum
  computing},\ }\href
  {https://doi.org/https://doi.org/10.1038/s41567-022-01613-z} {\bibfield
  {journal} {\bibinfo  {journal} {Nature Physics}\ }\textbf {\bibinfo {volume}
  {18}},\ \bibinfo {pages} {736} (\bibinfo {year} {2022})}\BibitemShut
  {NoStop}%
\bibitem [{\citenamefont {Yang}\ \emph {et~al.}(2024)\citenamefont {Yang},
  \citenamefont {Kladari\`c}, \citenamefont {Drimmer}, \citenamefont {von
  L\"upke}, \citenamefont {Lenterman}, \citenamefont {Bus}, \citenamefont
  {Marti}, \citenamefont {Fadel},\ and\ \citenamefont {Chu}}]{Yang2024}%
  \BibitemOpen
  \bibfield  {author} {\bibinfo {author} {\bibfnamefont {Y.}~\bibnamefont
  {Yang}}, \bibinfo {author} {\bibfnamefont {I.}~\bibnamefont {Kladari\`c}},
  \bibinfo {author} {\bibfnamefont {M.}~\bibnamefont {Drimmer}}, \bibinfo
  {author} {\bibfnamefont {U.}~\bibnamefont {von L\"upke}}, \bibinfo {author}
  {\bibfnamefont {D.}~\bibnamefont {Lenterman}}, \bibinfo {author}
  {\bibfnamefont {J.}~\bibnamefont {Bus}}, \bibinfo {author} {\bibfnamefont
  {S.}~\bibnamefont {Marti}}, \bibinfo {author} {\bibfnamefont
  {M.}~\bibnamefont {Fadel}},\ and\ \bibinfo {author} {\bibfnamefont
  {Y.}~\bibnamefont {Chu}},\ }\bibfield  {title} {\bibinfo {title} {A
  mechanical qubit},\ }\href {https://doi.org/10.1126/science.adr2464}
  {\bibfield  {journal} {\bibinfo  {journal} {Science}\ }\textbf {\bibinfo
  {volume} {386}},\ \bibinfo {pages} {783} (\bibinfo {year}
  {2024})}\BibitemShut {NoStop}%
\bibitem [{\citenamefont {Stannigel}\ \emph {et~al.}(2012)\citenamefont
  {Stannigel}, \citenamefont {Komar}, \citenamefont {Habraken}, \citenamefont
  {Bennett}, \citenamefont {Lukin}, \citenamefont {Zoller},\ and\ \citenamefont
  {Rabl}}]{Rabl2012}%
  \BibitemOpen
  \bibfield  {author} {\bibinfo {author} {\bibfnamefont {K.}~\bibnamefont
  {Stannigel}}, \bibinfo {author} {\bibfnamefont {P.}~\bibnamefont {Komar}},
  \bibinfo {author} {\bibfnamefont {S.~J.~M.}\ \bibnamefont {Habraken}},
  \bibinfo {author} {\bibfnamefont {S.~D.}\ \bibnamefont {Bennett}}, \bibinfo
  {author} {\bibfnamefont {M.~D.}\ \bibnamefont {Lukin}}, \bibinfo {author}
  {\bibfnamefont {P.}~\bibnamefont {Zoller}},\ and\ \bibinfo {author}
  {\bibfnamefont {P.}~\bibnamefont {Rabl}},\ }\bibfield  {title} {\bibinfo
  {title} {Optomechanical quantum information processing with photons and
  phonons},\ }\href {https://doi.org/10.1103/PhysRevLett.109.013603} {\bibfield
   {journal} {\bibinfo  {journal} {Phys. Rev. Lett.}\ }\textbf {\bibinfo
  {volume} {109}},\ \bibinfo {pages} {013603} (\bibinfo {year}
  {2012})}\BibitemShut {NoStop}%
\bibitem [{\citenamefont {Quan}\ \emph {et~al.}(2007)\citenamefont {Quan},
  \citenamefont {Liu}, \citenamefont {Sun},\ and\ \citenamefont
  {Nori}}]{Nori-QuantumHEatEng-2007}%
  \BibitemOpen
  \bibfield  {author} {\bibinfo {author} {\bibfnamefont {H.~T.}\ \bibnamefont
  {Quan}}, \bibinfo {author} {\bibfnamefont {Y.-X.}\ \bibnamefont {Liu}},
  \bibinfo {author} {\bibfnamefont {C.~P.}\ \bibnamefont {Sun}},\ and\ \bibinfo
  {author} {\bibfnamefont {F.}~\bibnamefont {Nori}},\ }\bibfield  {title}
  {\bibinfo {title} {Quantum thermodynamic cycles and quantum heat engines},\
  }\href {https://doi.org/10.1103/PhysRevE.76.031105} {\bibfield  {journal}
  {\bibinfo  {journal} {Phys. Rev. E}\ }\textbf {\bibinfo {volume} {76}},\
  \bibinfo {pages} {031105} (\bibinfo {year} {2007})}\BibitemShut {NoStop}%
\bibitem [{\citenamefont {Dong}\ \emph {et~al.}(2015)\citenamefont {Dong},
  \citenamefont {Zhang}, \citenamefont {Bariani},\ and\ \citenamefont
  {Meystre}}]{Bariani-QuantumHEatEng-I}%
  \BibitemOpen
  \bibfield  {author} {\bibinfo {author} {\bibfnamefont {Y.}~\bibnamefont
  {Dong}}, \bibinfo {author} {\bibfnamefont {K.}~\bibnamefont {Zhang}},
  \bibinfo {author} {\bibfnamefont {F.}~\bibnamefont {Bariani}},\ and\ \bibinfo
  {author} {\bibfnamefont {P.}~\bibnamefont {Meystre}},\ }\bibfield  {title}
  {\bibinfo {title} {Work measurement in an optomechanical quantum heat
  engine},\ }\href {https://doi.org/10.1103/PhysRevA.92.033854} {\bibfield
  {journal} {\bibinfo  {journal} {Phys. Rev. A}\ }\textbf {\bibinfo {volume}
  {92}},\ \bibinfo {pages} {033854} (\bibinfo {year} {2015})}\BibitemShut
  {NoStop}%
\bibitem [{\citenamefont {Zhang}\ \emph
  {et~al.}(2014{\natexlab{a}})\citenamefont {Zhang}, \citenamefont {Bariani},\
  and\ \citenamefont {Meystre}}]{Bariani-QuantumHEatEng-II}%
  \BibitemOpen
  \bibfield  {author} {\bibinfo {author} {\bibfnamefont {K.}~\bibnamefont
  {Zhang}}, \bibinfo {author} {\bibfnamefont {F.}~\bibnamefont {Bariani}},\
  and\ \bibinfo {author} {\bibfnamefont {P.}~\bibnamefont {Meystre}},\
  }\bibfield  {title} {\bibinfo {title} {Theory of an optomechanical quantum
  heat engine},\ }\href {https://doi.org/10.1103/PhysRevA.90.023819} {\bibfield
   {journal} {\bibinfo  {journal} {Phys. Rev. A}\ }\textbf {\bibinfo {volume}
  {90}},\ \bibinfo {pages} {023819} (\bibinfo {year}
  {2014}{\natexlab{a}})}\BibitemShut {NoStop}%
\bibitem [{\citenamefont {Zhang}\ \emph
  {et~al.}(2014{\natexlab{b}})\citenamefont {Zhang}, \citenamefont {Bariani},\
  and\ \citenamefont {Meystre}}]{Bariani-QuantumHEatEng-III}%
  \BibitemOpen
  \bibfield  {author} {\bibinfo {author} {\bibfnamefont {K.}~\bibnamefont
  {Zhang}}, \bibinfo {author} {\bibfnamefont {F.}~\bibnamefont {Bariani}},\
  and\ \bibinfo {author} {\bibfnamefont {P.}~\bibnamefont {Meystre}},\
  }\bibfield  {title} {\bibinfo {title} {Quantum optomechanical heat engine},\
  }\href {https://doi.org/10.1103/PhysRevLett.112.150602} {\bibfield  {journal}
  {\bibinfo  {journal} {Phys. Rev. Lett.}\ }\textbf {\bibinfo {volume} {112}},\
  \bibinfo {pages} {150602} (\bibinfo {year} {2014}{\natexlab{b}})}\BibitemShut
  {NoStop}%
\bibitem [{\citenamefont {Ferreri}\ \emph {et~al.}(2023)\citenamefont
  {Ferreri}, \citenamefont {Macr{\`\i}}, \citenamefont {Wilhelm}, \citenamefont
  {Nori},\ and\ \citenamefont {Bruschi}}]{Nori-QuantumHEatEng-2023}%
  \BibitemOpen
  \bibfield  {author} {\bibinfo {author} {\bibfnamefont {A.}~\bibnamefont
  {Ferreri}}, \bibinfo {author} {\bibfnamefont {V.}~\bibnamefont {Macr{\`\i}}},
  \bibinfo {author} {\bibfnamefont {F.~K.}\ \bibnamefont {Wilhelm}}, \bibinfo
  {author} {\bibfnamefont {F.}~\bibnamefont {Nori}},\ and\ \bibinfo {author}
  {\bibfnamefont {D.~E.}\ \bibnamefont {Bruschi}},\ }\bibfield  {title}
  {\bibinfo {title} {Quantum field heat engine powered by phonon-photon
  interactions},\ }\href@noop {} {\bibfield  {journal} {\bibinfo  {journal}
  {arXiv preprint arXiv:2305.06445}\ } (\bibinfo {year} {2023})}\BibitemShut
  {NoStop}%
\bibitem [{\citenamefont {Law}(1995)}]{Law1995}%
  \BibitemOpen
  \bibfield  {author} {\bibinfo {author} {\bibfnamefont {C.~K.}\ \bibnamefont
  {Law}},\ }\bibfield  {title} {\bibinfo {title} {Interaction between a moving
  mirror and radiation pressure: A hamiltonian formulation},\ }\href
  {https://doi.org/10.1103/PhysRevA.51.2537} {\bibfield  {journal} {\bibinfo
  {journal} {Phys. Rev. A}\ }\textbf {\bibinfo {volume} {51}},\ \bibinfo
  {pages} {2537} (\bibinfo {year} {1995})}\BibitemShut {NoStop}%
\bibitem [{\citenamefont {Kippenberg}\ and\ \citenamefont
  {Vahala}(2008)}]{kippenberg2008cavity}%
  \BibitemOpen
  \bibfield  {author} {\bibinfo {author} {\bibfnamefont {T.~J.}\ \bibnamefont
  {Kippenberg}}\ and\ \bibinfo {author} {\bibfnamefont {K.~J.}\ \bibnamefont
  {Vahala}},\ }\bibfield  {title} {\bibinfo {title} {Cavity optomechanics:
  back-action at the mesoscale},\ }\href
  {https://doi.org/https://doi.org/10.1126/science.1156032} {\bibfield
  {journal} {\bibinfo  {journal} {Science}\ }\textbf {\bibinfo {volume}
  {321}},\ \bibinfo {pages} {1172} (\bibinfo {year} {2008})}\BibitemShut
  {NoStop}%
\bibitem [{\citenamefont {O’Connell}\ \emph
  {et~al.}(2010{\natexlab{b}})\citenamefont {O’Connell}, \citenamefont
  {Hofheinz}, \citenamefont {Ansmann}, \citenamefont {Bialczak}, \citenamefont
  {Lenander}, \citenamefont {Lucero}, \citenamefont {Neeley}, \citenamefont
  {Sank}, \citenamefont {Wang}, \citenamefont {Weides} \emph
  {et~al.}}]{Cleland2010}%
  \BibitemOpen
  \bibfield  {author} {\bibinfo {author} {\bibfnamefont {A.~D.}\ \bibnamefont
  {O’Connell}}, \bibinfo {author} {\bibfnamefont {M.}~\bibnamefont
  {Hofheinz}}, \bibinfo {author} {\bibfnamefont {M.}~\bibnamefont {Ansmann}},
  \bibinfo {author} {\bibfnamefont {R.~C.}\ \bibnamefont {Bialczak}}, \bibinfo
  {author} {\bibfnamefont {M.}~\bibnamefont {Lenander}}, \bibinfo {author}
  {\bibfnamefont {E.}~\bibnamefont {Lucero}}, \bibinfo {author} {\bibfnamefont
  {M.}~\bibnamefont {Neeley}}, \bibinfo {author} {\bibfnamefont
  {D.}~\bibnamefont {Sank}}, \bibinfo {author} {\bibfnamefont {H.}~\bibnamefont
  {Wang}}, \bibinfo {author} {\bibfnamefont {M.}~\bibnamefont {Weides}}, \emph
  {et~al.},\ }\bibfield  {title} {\bibinfo {title} {Quantum ground state and
  single-phonon control of a mechanical resonator},\ }\href@noop {} {\bibfield
  {journal} {\bibinfo  {journal} {Nature}\ }\textbf {\bibinfo {volume} {464}},\
  \bibinfo {pages} {697} (\bibinfo {year} {2010}{\natexlab{b}})}\BibitemShut
  {NoStop}%
\bibitem [{\citenamefont {Nunnenkamp}\ \emph {et~al.}(2011)\citenamefont
  {Nunnenkamp}, \citenamefont {B\o{}rkje},\ and\ \citenamefont
  {Girvin}}]{Girvin2011_SinglePhoton}%
  \BibitemOpen
  \bibfield  {author} {\bibinfo {author} {\bibfnamefont {A.}~\bibnamefont
  {Nunnenkamp}}, \bibinfo {author} {\bibfnamefont {K.}~\bibnamefont
  {B\o{}rkje}},\ and\ \bibinfo {author} {\bibfnamefont {S.~M.}\ \bibnamefont
  {Girvin}},\ }\bibfield  {title} {\bibinfo {title} {Single-photon
  optomechanics},\ }\href {https://doi.org/10.1103/PhysRevLett.107.063602}
  {\bibfield  {journal} {\bibinfo  {journal} {Phys. Rev. Lett.}\ }\textbf
  {\bibinfo {volume} {107}},\ \bibinfo {pages} {063602} (\bibinfo {year}
  {2011})}\BibitemShut {NoStop}%
\bibitem [{\citenamefont {Regal}\ \emph {et~al.}(2008)\citenamefont {Regal},
  \citenamefont {Teufel},\ and\ \citenamefont {Lehnert}}]{Regal2008}%
  \BibitemOpen
  \bibfield  {author} {\bibinfo {author} {\bibfnamefont {C.~A.}\ \bibnamefont
  {Regal}}, \bibinfo {author} {\bibfnamefont {J.~D.}\ \bibnamefont {Teufel}},\
  and\ \bibinfo {author} {\bibfnamefont {K.~W.}\ \bibnamefont {Lehnert}},\
  }\bibfield  {title} {\bibinfo {title} {Measuring nanomechanical motion with a
  microwave cavity interferometer},\ }\href {https://doi.org/10.1038/nphys974}
  {\bibfield  {journal} {\bibinfo  {journal} {Nat. Physics}\ }\textbf {\bibinfo
  {volume} {4}},\ \bibinfo {pages} {555} (\bibinfo {year} {2008})}\BibitemShut
  {NoStop}%
\bibitem [{\citenamefont {LaHaye}\ \emph {et~al.}(2009)\citenamefont {LaHaye},
  \citenamefont {Suh}, \citenamefont {Echternach}, \citenamefont {Schwab},\
  and\ \citenamefont {Roukes}}]{lahaye2009nanomechanical}%
  \BibitemOpen
  \bibfield  {author} {\bibinfo {author} {\bibfnamefont {M.}~\bibnamefont
  {LaHaye}}, \bibinfo {author} {\bibfnamefont {J.}~\bibnamefont {Suh}},
  \bibinfo {author} {\bibfnamefont {P.}~\bibnamefont {Echternach}}, \bibinfo
  {author} {\bibfnamefont {K.~C.}\ \bibnamefont {Schwab}},\ and\ \bibinfo
  {author} {\bibfnamefont {M.~L.}\ \bibnamefont {Roukes}},\ }\bibfield  {title}
  {\bibinfo {title} {Nanomechanical measurements of a superconducting qubit},\
  }\href {https://doi.org/10.1038/nature08093} {\bibfield  {journal} {\bibinfo
  {journal} {Nature}\ }\textbf {\bibinfo {volume} {459}},\ \bibinfo {pages}
  {960} (\bibinfo {year} {2009})}\BibitemShut {NoStop}%
\bibitem [{\citenamefont {Teufel}\ \emph
  {et~al.}(2011{\natexlab{b}})\citenamefont {Teufel}, \citenamefont {Li},
  \citenamefont {Allman}, \citenamefont {Cicak}, \citenamefont {Sirois},
  \citenamefont {Whittaker},\ and\ \citenamefont
  {Simmonds}}]{teufel2011circuit}%
  \BibitemOpen
  \bibfield  {author} {\bibinfo {author} {\bibfnamefont {J.~D.}\ \bibnamefont
  {Teufel}}, \bibinfo {author} {\bibfnamefont {D.}~\bibnamefont {Li}}, \bibinfo
  {author} {\bibfnamefont {M.~S.}\ \bibnamefont {Allman}}, \bibinfo {author}
  {\bibfnamefont {K.}~\bibnamefont {Cicak}}, \bibinfo {author} {\bibfnamefont
  {A.~J.}\ \bibnamefont {Sirois}}, \bibinfo {author} {\bibfnamefont {J.~D.}\
  \bibnamefont {Whittaker}},\ and\ \bibinfo {author} {\bibfnamefont {R.~W.}\
  \bibnamefont {Simmonds}},\ }\bibfield  {title} {\bibinfo {title} {Circuit
  cavity electromechanics in the strong-coupling regime},\ }\href
  {https://doi.org/10.1038/nature09898} {\bibfield  {journal} {\bibinfo
  {journal} {Nature}\ }\textbf {\bibinfo {volume} {471}},\ \bibinfo {pages}
  {204} (\bibinfo {year} {2011}{\natexlab{b}})}\BibitemShut {NoStop}%
\bibitem [{\citenamefont {Forn-D\'{\i}az}\ \emph {et~al.}(2019)\citenamefont
  {Forn-D\'{\i}az}, \citenamefont {Lamata}, \citenamefont {Rico}, \citenamefont
  {Kono},\ and\ \citenamefont {Solano}}]{UltraStrong_Review}%
  \BibitemOpen
  \bibfield  {author} {\bibinfo {author} {\bibfnamefont {P.}~\bibnamefont
  {Forn-D\'{\i}az}}, \bibinfo {author} {\bibfnamefont {L.}~\bibnamefont
  {Lamata}}, \bibinfo {author} {\bibfnamefont {E.}~\bibnamefont {Rico}},
  \bibinfo {author} {\bibfnamefont {J.}~\bibnamefont {Kono}},\ and\ \bibinfo
  {author} {\bibfnamefont {E.}~\bibnamefont {Solano}},\ }\bibfield  {title}
  {\bibinfo {title} {Ultrastrong coupling regimes of light-matter
  interaction},\ }\href {https://doi.org/10.1103/RevModPhys.91.025005}
  {\bibfield  {journal} {\bibinfo  {journal} {Rev. Mod. Phys.}\ }\textbf
  {\bibinfo {volume} {91}},\ \bibinfo {pages} {025005} (\bibinfo {year}
  {2019})}\BibitemShut {NoStop}%
\bibitem [{\citenamefont {Rouxinol}\ \emph {et~al.}(2016)\citenamefont
  {Rouxinol}, \citenamefont {Hao}, \citenamefont {Brito}, \citenamefont
  {Caldeira}, \citenamefont {Irish},\ and\ \citenamefont
  {LaHaye}}]{Rouxinol-2016}%
  \BibitemOpen
  \bibfield  {author} {\bibinfo {author} {\bibfnamefont {F.}~\bibnamefont
  {Rouxinol}}, \bibinfo {author} {\bibfnamefont {Y.}~\bibnamefont {Hao}},
  \bibinfo {author} {\bibfnamefont {F.}~\bibnamefont {Brito}}, \bibinfo
  {author} {\bibfnamefont {A.}~\bibnamefont {Caldeira}}, \bibinfo {author}
  {\bibfnamefont {E.}~\bibnamefont {Irish}},\ and\ \bibinfo {author}
  {\bibfnamefont {M.}~\bibnamefont {LaHaye}},\ }\bibfield  {title} {\bibinfo
  {title} {Measurements of nanoresonator-qubit interactions in a hybrid quantum
  electromechanical system},\ }\href
  {https://doi.org/10.1088/0957-4484/27/36/364003} {\bibfield  {journal}
  {\bibinfo  {journal} {Nanotechnology}\ }\textbf {\bibinfo {volume} {27}},\
  \bibinfo {pages} {364003} (\bibinfo {year} {2016})}\BibitemShut {NoStop}%
\bibitem [{\citenamefont {Heikkil\"a}\ \emph {et~al.}(2014)\citenamefont
  {Heikkil\"a}, \citenamefont {Massel}, \citenamefont {Tuorila}, \citenamefont
  {Khan},\ and\ \citenamefont {Sillanp\"a\"a}}]{Sillanpaa2014}%
  \BibitemOpen
  \bibfield  {author} {\bibinfo {author} {\bibfnamefont {T.~T.}\ \bibnamefont
  {Heikkil\"a}}, \bibinfo {author} {\bibfnamefont {F.}~\bibnamefont {Massel}},
  \bibinfo {author} {\bibfnamefont {J.}~\bibnamefont {Tuorila}}, \bibinfo
  {author} {\bibfnamefont {R.}~\bibnamefont {Khan}},\ and\ \bibinfo {author}
  {\bibfnamefont {M.~A.}\ \bibnamefont {Sillanp\"a\"a}},\ }\bibfield  {title}
  {\bibinfo {title} {Enhancing optomechanical coupling via the josephson
  effect},\ }\href {https://doi.org/10.1103/PhysRevLett.112.203603} {\bibfield
  {journal} {\bibinfo  {journal} {Phys. Rev. Lett.}\ }\textbf {\bibinfo
  {volume} {112}},\ \bibinfo {pages} {203603} (\bibinfo {year}
  {2014})}\BibitemShut {NoStop}%
\bibitem [{\citenamefont {Pirkkalainen}\ \emph {et~al.}(2015)\citenamefont
  {Pirkkalainen}, \citenamefont {Cho}, \citenamefont {Massel}, \citenamefont
  {Tuorila}, \citenamefont {Heikkil{\"a}}, \citenamefont {Hakonen},\ and\
  \citenamefont {Sillanp{\"a}{\"a}}}]{pirkkalainen2015}%
  \BibitemOpen
  \bibfield  {author} {\bibinfo {author} {\bibfnamefont {J.-M.}\ \bibnamefont
  {Pirkkalainen}}, \bibinfo {author} {\bibfnamefont {S.}~\bibnamefont {Cho}},
  \bibinfo {author} {\bibfnamefont {F.}~\bibnamefont {Massel}}, \bibinfo
  {author} {\bibfnamefont {J.}~\bibnamefont {Tuorila}}, \bibinfo {author}
  {\bibfnamefont {T.}~\bibnamefont {Heikkil{\"a}}}, \bibinfo {author}
  {\bibfnamefont {P.}~\bibnamefont {Hakonen}},\ and\ \bibinfo {author}
  {\bibfnamefont {M.}~\bibnamefont {Sillanp{\"a}{\"a}}},\ }\bibfield  {title}
  {\bibinfo {title} {Cavity optomechanics mediated by a quantum two-level
  system},\ }\href@noop {} {\bibfield  {journal} {\bibinfo  {journal} {Nature
  communications}\ }\textbf {\bibinfo {volume} {6}},\ \bibinfo {pages} {6981}
  (\bibinfo {year} {2015})}\BibitemShut {NoStop}%
\bibitem [{\citenamefont {Aporvari}\ and\ \citenamefont
  {Vitali}(2021)}]{Vitali2021}%
  \BibitemOpen
  \bibfield  {author} {\bibinfo {author} {\bibfnamefont {A.~S.}\ \bibnamefont
  {Aporvari}}\ and\ \bibinfo {author} {\bibfnamefont {D.}~\bibnamefont
  {Vitali}},\ }\bibfield  {title} {\bibinfo {title} {Strong coupling
  optomechanics mediated by a qubit in the dispersive regime},\ }\bibfield
  {journal} {\bibinfo  {journal} {Entropy}\ }\textbf {\bibinfo {volume} {23}},\
  \href {https://doi.org/10.3390/e23080966} {10.3390/e23080966} (\bibinfo
  {year} {2021})\BibitemShut {NoStop}%
\bibitem [{\citenamefont {Manninen}\ \emph {et~al.}(2022)\citenamefont
  {Manninen}, \citenamefont {Haque}, \citenamefont {Vitali},\ and\
  \citenamefont {Hakonen}}]{Hakonen-PRB-2022}%
  \BibitemOpen
  \bibfield  {author} {\bibinfo {author} {\bibfnamefont {J.}~\bibnamefont
  {Manninen}}, \bibinfo {author} {\bibfnamefont {M.~T.}\ \bibnamefont {Haque}},
  \bibinfo {author} {\bibfnamefont {D.}~\bibnamefont {Vitali}},\ and\ \bibinfo
  {author} {\bibfnamefont {P.}~\bibnamefont {Hakonen}},\ }\bibfield  {title}
  {\bibinfo {title} {Enhancement of the optomechanical coupling and kerr
  nonlinearity using the josephson capacitance of a cooper-pair box},\ }\href
  {https://doi.org/10.1103/PhysRevB.105.144508} {\bibfield  {journal} {\bibinfo
   {journal} {Phys. Rev. B}\ }\textbf {\bibinfo {volume} {105}},\ \bibinfo
  {pages} {144508} (\bibinfo {year} {2022})}\BibitemShut {NoStop}%
\bibitem [{\citenamefont {Johansson}\ \emph {et~al.}(2009)\citenamefont
  {Johansson}, \citenamefont {Johansson}, \citenamefont {Wilson},\ and\
  \citenamefont {Nori}}]{Johansson-PRL-2009}%
  \BibitemOpen
  \bibfield  {author} {\bibinfo {author} {\bibfnamefont {J.~R.}\ \bibnamefont
  {Johansson}}, \bibinfo {author} {\bibfnamefont {G.}~\bibnamefont
  {Johansson}}, \bibinfo {author} {\bibfnamefont {C.~M.}\ \bibnamefont
  {Wilson}},\ and\ \bibinfo {author} {\bibfnamefont {F.}~\bibnamefont {Nori}},\
  }\bibfield  {title} {\bibinfo {title} {Dynamical casimir effect in a
  superconducting coplanar waveguide},\ }\href
  {https://doi.org/10.1103/PhysRevLett.103.147003} {\bibfield  {journal}
  {\bibinfo  {journal} {Phys. Rev. Lett.}\ }\textbf {\bibinfo {volume} {103}},\
  \bibinfo {pages} {147003} (\bibinfo {year} {2009})}\BibitemShut {NoStop}%
\bibitem [{\citenamefont {Johansson}\ \emph {et~al.}(2010)\citenamefont
  {Johansson}, \citenamefont {Johansson}, \citenamefont {Wilson},\ and\
  \citenamefont {Nori}}]{Johansson-PRA-2010}%
  \BibitemOpen
  \bibfield  {author} {\bibinfo {author} {\bibfnamefont {J.~R.}\ \bibnamefont
  {Johansson}}, \bibinfo {author} {\bibfnamefont {G.}~\bibnamefont
  {Johansson}}, \bibinfo {author} {\bibfnamefont {C.~M.}\ \bibnamefont
  {Wilson}},\ and\ \bibinfo {author} {\bibfnamefont {F.}~\bibnamefont {Nori}},\
  }\bibfield  {title} {\bibinfo {title} {Dynamical casimir effect in
  superconducting microwave circuits},\ }\href
  {https://doi.org/10.1103/PhysRevA.82.052509} {\bibfield  {journal} {\bibinfo
  {journal} {Phys. Rev. A}\ }\textbf {\bibinfo {volume} {82}},\ \bibinfo
  {pages} {052509} (\bibinfo {year} {2010})}\BibitemShut {NoStop}%
\bibitem [{\citenamefont {Wilson}\ \emph {et~al.}(2011)\citenamefont {Wilson},
  \citenamefont {Johansson}, \citenamefont {Pourkabirian}, \citenamefont
  {Simoen}, \citenamefont {Johansson}, \citenamefont {Duty}, \citenamefont
  {Nori},\ and\ \citenamefont {Delsing}}]{Wilson-DCE-Analog-2011}%
  \BibitemOpen
  \bibfield  {author} {\bibinfo {author} {\bibfnamefont {C.~W.}\ \bibnamefont
  {Wilson}}, \bibinfo {author} {\bibfnamefont {G.}~\bibnamefont {Johansson}},
  \bibinfo {author} {\bibfnamefont {A.}~\bibnamefont {Pourkabirian}}, \bibinfo
  {author} {\bibfnamefont {M.}~\bibnamefont {Simoen}}, \bibinfo {author}
  {\bibfnamefont {J.~R.}\ \bibnamefont {Johansson}}, \bibinfo {author}
  {\bibfnamefont {T.}~\bibnamefont {Duty}}, \bibinfo {author} {\bibfnamefont
  {F.}~\bibnamefont {Nori}},\ and\ \bibinfo {author} {\bibfnamefont
  {P.}~\bibnamefont {Delsing}},\ }\bibfield  {title} {\bibinfo {title}
  {Observation of the dynamical casimir effect in a superconducting circuit},\
  }\href {https://doi.org/10.1038/nature10561} {\bibfield  {journal} {\bibinfo
  {journal} {Nature}\ }\textbf {\bibinfo {volume} {479}},\ \bibinfo {pages}
  {376} (\bibinfo {year} {2011})}\BibitemShut {NoStop}%
\bibitem [{\citenamefont {Butera}\ and\ \citenamefont
  {Carusotto}(2019)}]{Butera-PRA-2019}%
  \BibitemOpen
  \bibfield  {author} {\bibinfo {author} {\bibfnamefont {S.}~\bibnamefont
  {Butera}}\ and\ \bibinfo {author} {\bibfnamefont {I.}~\bibnamefont
  {Carusotto}},\ }\bibfield  {title} {\bibinfo {title} {Mechanical backreaction
  effect of the dynamical casimir emission},\ }\href
  {https://doi.org/10.1103/PhysRevA.99.053815} {\bibfield  {journal} {\bibinfo
  {journal} {Phys. Rev. A}\ }\textbf {\bibinfo {volume} {99}},\ \bibinfo
  {pages} {053815} (\bibinfo {year} {2019})}\BibitemShut {NoStop}%
\bibitem [{\citenamefont {Bose}\ \emph {et~al.}(1997)\citenamefont {Bose},
  \citenamefont {Jacobs},\ and\ \citenamefont {Knight}}]{Knight1997}%
  \BibitemOpen
  \bibfield  {author} {\bibinfo {author} {\bibfnamefont {S.}~\bibnamefont
  {Bose}}, \bibinfo {author} {\bibfnamefont {K.}~\bibnamefont {Jacobs}},\ and\
  \bibinfo {author} {\bibfnamefont {P.~L.}\ \bibnamefont {Knight}},\ }\bibfield
   {title} {\bibinfo {title} {Preparation of nonclassical states in cavities
  with a moving mirror},\ }\href {https://doi.org/10.1103/PhysRevA.56.4175}
  {\bibfield  {journal} {\bibinfo  {journal} {Phys. Rev. A}\ }\textbf {\bibinfo
  {volume} {56}},\ \bibinfo {pages} {4175} (\bibinfo {year}
  {1997})}\BibitemShut {NoStop}%
\bibitem [{\citenamefont {Mancini}\ \emph {et~al.}(1997)\citenamefont
  {Mancini}, \citenamefont {Man'ko},\ and\ \citenamefont
  {Tombesi}}]{Tombesi1997}%
  \BibitemOpen
  \bibfield  {author} {\bibinfo {author} {\bibfnamefont {S.}~\bibnamefont
  {Mancini}}, \bibinfo {author} {\bibfnamefont {V.~I.}\ \bibnamefont
  {Man'ko}},\ and\ \bibinfo {author} {\bibfnamefont {P.}~\bibnamefont
  {Tombesi}},\ }\bibfield  {title} {\bibinfo {title} {Ponderomotive control of
  quantum macroscopic coherence},\ }\href
  {https://doi.org/10.1103/PhysRevA.55.3042} {\bibfield  {journal} {\bibinfo
  {journal} {Phys. Rev. A}\ }\textbf {\bibinfo {volume} {55}},\ \bibinfo
  {pages} {3042} (\bibinfo {year} {1997})}\BibitemShut {NoStop}%
\bibitem [{\citenamefont {Garziano}\ \emph {et~al.}(2015)\citenamefont
  {Garziano}, \citenamefont {Stassi}, \citenamefont {Macr\'{\i}}, \citenamefont
  {Savasta},\ and\ \citenamefont {Di~Stefano}}]{Savasta2015}%
  \BibitemOpen
  \bibfield  {author} {\bibinfo {author} {\bibfnamefont {L.}~\bibnamefont
  {Garziano}}, \bibinfo {author} {\bibfnamefont {R.}~\bibnamefont {Stassi}},
  \bibinfo {author} {\bibfnamefont {V.}~\bibnamefont {Macr\'{\i}}}, \bibinfo
  {author} {\bibfnamefont {S.}~\bibnamefont {Savasta}},\ and\ \bibinfo {author}
  {\bibfnamefont {O.}~\bibnamefont {Di~Stefano}},\ }\bibfield  {title}
  {\bibinfo {title} {Single-step arbitrary control of mechanical quantum states
  in ultrastrong optomechanics},\ }\href
  {https://doi.org/10.1103/PhysRevA.91.023809} {\bibfield  {journal} {\bibinfo
  {journal} {Phys. Rev. A}\ }\textbf {\bibinfo {volume} {91}},\ \bibinfo
  {pages} {023809} (\bibinfo {year} {2015})}\BibitemShut {NoStop}%
\bibitem [{\citenamefont {Butera}\ and\ \citenamefont
  {Passante}(2013)}]{Giulio-PRL-2013}%
  \BibitemOpen
  \bibfield  {author} {\bibinfo {author} {\bibfnamefont {S.}~\bibnamefont
  {Butera}}\ and\ \bibinfo {author} {\bibfnamefont {R.}~\bibnamefont
  {Passante}},\ }\bibfield  {title} {\bibinfo {title} {Field fluctuations in a
  one-dimensional cavity with a mobile wall},\ }\href
  {https://doi.org/10.1103/PhysRevLett.111.060403} {\bibfield  {journal}
  {\bibinfo  {journal} {Phys. Rev. Lett.}\ }\textbf {\bibinfo {volume} {111}},\
  \bibinfo {pages} {060403} (\bibinfo {year} {2013})}\BibitemShut {NoStop}%
\bibitem [{\citenamefont {Armata}\ and\ \citenamefont
  {Passante}(2015)}]{Armata-PRD-2015}%
  \BibitemOpen
  \bibfield  {author} {\bibinfo {author} {\bibfnamefont {F.}~\bibnamefont
  {Armata}}\ and\ \bibinfo {author} {\bibfnamefont {R.}~\bibnamefont
  {Passante}},\ }\bibfield  {title} {\bibinfo {title} {Vacuum energy densities
  of a field in a cavity with a mobile boundary},\ }\href
  {https://doi.org/10.1103/PhysRevD.91.025012} {\bibfield  {journal} {\bibinfo
  {journal} {Phys. Rev. D}\ }\textbf {\bibinfo {volume} {91}},\ \bibinfo
  {pages} {025012} (\bibinfo {year} {2015})}\BibitemShut {NoStop}%
\bibitem [{\citenamefont {Armata}\ \emph {et~al.}(2017)\citenamefont {Armata},
  \citenamefont {Kim}, \citenamefont {Butera}, \citenamefont {Rizzuto},\ and\
  \citenamefont {Passante}}]{Armata-PRD-2017}%
  \BibitemOpen
  \bibfield  {author} {\bibinfo {author} {\bibfnamefont {F.}~\bibnamefont
  {Armata}}, \bibinfo {author} {\bibfnamefont {M.}~\bibnamefont {Kim}},
  \bibinfo {author} {\bibfnamefont {S.}~\bibnamefont {Butera}}, \bibinfo
  {author} {\bibfnamefont {L.}~\bibnamefont {Rizzuto}},\ and\ \bibinfo {author}
  {\bibfnamefont {R.}~\bibnamefont {Passante}},\ }\bibfield  {title} {\bibinfo
  {title} {Nonequilibrium dressing in a cavity with a movable reflecting
  mirror},\ }\href {https://doi.org/10.1103/PhysRevD.96.045007} {\bibfield
  {journal} {\bibinfo  {journal} {Phys. Rev. D}\ }\textbf {\bibinfo {volume}
  {96}},\ \bibinfo {pages} {045007} (\bibinfo {year} {2017})}\BibitemShut
  {NoStop}%
\bibitem [{\citenamefont {Kardar}\ and\ \citenamefont
  {Golestanian}(1999)}]{KardarRMP1999}%
  \BibitemOpen
  \bibfield  {author} {\bibinfo {author} {\bibfnamefont {M.}~\bibnamefont
  {Kardar}}\ and\ \bibinfo {author} {\bibfnamefont {R.}~\bibnamefont
  {Golestanian}},\ }\bibfield  {title} {\bibinfo {title} {The ``friction'' of
  vacuum, and other fluctuation-induced forces},\ }\href
  {https://doi.org/10.1103/RevModPhys.71.1233} {\bibfield  {journal} {\bibinfo
  {journal} {Rev. Mod. Phys.}\ }\textbf {\bibinfo {volume} {71}},\ \bibinfo
  {pages} {1233} (\bibinfo {year} {1999})}\BibitemShut {NoStop}%
\bibitem [{\citenamefont {Macr\`{\i}}\ \emph {et~al.}(2018)\citenamefont
  {Macr\`{\i}}, \citenamefont {Ridolfo}, \citenamefont {Di~Stefano},
  \citenamefont {Kockum}, \citenamefont {Nori},\ and\ \citenamefont
  {Savasta}}]{Savasta-PRX-2018}%
  \BibitemOpen
  \bibfield  {author} {\bibinfo {author} {\bibfnamefont {V.}~\bibnamefont
  {Macr\`{\i}}}, \bibinfo {author} {\bibfnamefont {A.}~\bibnamefont {Ridolfo}},
  \bibinfo {author} {\bibfnamefont {O.}~\bibnamefont {Di~Stefano}}, \bibinfo
  {author} {\bibfnamefont {A.~F.}\ \bibnamefont {Kockum}}, \bibinfo {author}
  {\bibfnamefont {F.}~\bibnamefont {Nori}},\ and\ \bibinfo {author}
  {\bibfnamefont {S.}~\bibnamefont {Savasta}},\ }\bibfield  {title} {\bibinfo
  {title} {{Nonperturbative Dynamical Casimir Effect in Optomechanical Systems:
  Vacuum Casimir-Rabi Splittings}},\ }\href
  {https://doi.org/10.1103/PhysRevX.8.011031} {\bibfield  {journal} {\bibinfo
  {journal} {Phys. Rev. X}\ }\textbf {\bibinfo {volume} {8}},\ \bibinfo {pages}
  {011031} (\bibinfo {year} {2018})}\BibitemShut {NoStop}%
\bibitem [{\citenamefont {Di~Stefano}\ \emph {et~al.}(2019)\citenamefont
  {Di~Stefano}, \citenamefont {Settineri}, \citenamefont {Macr\`{\i}},
  \citenamefont {Ridolfo}, \citenamefont {Stassi}, \citenamefont {Kockum},
  \citenamefont {Savasta},\ and\ \citenamefont {Nori}}]{Savasta-PRL-2019}%
  \BibitemOpen
  \bibfield  {author} {\bibinfo {author} {\bibfnamefont {O.}~\bibnamefont
  {Di~Stefano}}, \bibinfo {author} {\bibfnamefont {A.}~\bibnamefont
  {Settineri}}, \bibinfo {author} {\bibfnamefont {V.}~\bibnamefont
  {Macr\`{\i}}}, \bibinfo {author} {\bibfnamefont {A.}~\bibnamefont {Ridolfo}},
  \bibinfo {author} {\bibfnamefont {R.}~\bibnamefont {Stassi}}, \bibinfo
  {author} {\bibfnamefont {A.}~\bibnamefont {Kockum}}, \bibinfo {author}
  {\bibfnamefont {S.}~\bibnamefont {Savasta}},\ and\ \bibinfo {author}
  {\bibfnamefont {F.}~\bibnamefont {Nori}},\ }\bibfield  {title} {\bibinfo
  {title} {Interaction of mechanical oscillators mediated by the exchange of
  virtual photon pairs},\ }\href
  {https://doi.org/10.1103/PhysRevLett.122.030402} {\bibfield  {journal}
  {\bibinfo  {journal} {Phys. Rev. Lett.}\ }\textbf {\bibinfo {volume} {122}},\
  \bibinfo {pages} {030402} (\bibinfo {year} {2019})}\BibitemShut {NoStop}%
\bibitem [{\citenamefont {Butera}\ and\ \citenamefont
  {Carusotto}(2020)}]{Butera-EPL-2019}%
  \BibitemOpen
  \bibfield  {author} {\bibinfo {author} {\bibfnamefont {S.}~\bibnamefont
  {Butera}}\ and\ \bibinfo {author} {\bibfnamefont {I.}~\bibnamefont
  {Carusotto}},\ }\bibfield  {title} {\bibinfo {title} {Quantum fluctuations of
  the friction force induced by the dynamical casimir emission},\ }\href
  {https://doi.org/10.1209/0295-5075/128/24002} {\bibfield  {journal} {\bibinfo
   {journal} {EPL (Europhysics Letters)}\ }\textbf {\bibinfo {volume} {128}},\
  \bibinfo {pages} {24002} (\bibinfo {year} {2020})}\BibitemShut {NoStop}%
\bibitem [{\citenamefont {Butera}(2023)}]{butera2023noise}%
  \BibitemOpen
  \bibfield  {author} {\bibinfo {author} {\bibfnamefont {S.}~\bibnamefont
  {Butera}},\ }\bibfield  {title} {\bibinfo {title} {Noise and dissipation on a
  moving mirror induced by the dynamical casimir emission},\ }\href
  {https://doi.org/10.1088/2515-7647/acff56} {\bibfield  {journal} {\bibinfo
  {journal} {Journal of Physics: Photonics}\ }\textbf {\bibinfo {volume} {5}},\
  \bibinfo {pages} {045003} (\bibinfo {year} {2023})}\BibitemShut {NoStop}%
\bibitem [{\citenamefont {Dalvit}\ and\ \citenamefont
  {Maia~Neto}(2000)}]{Dalvit-PRL-2000}%
  \BibitemOpen
  \bibfield  {author} {\bibinfo {author} {\bibfnamefont {D.~A.~R.}\
  \bibnamefont {Dalvit}}\ and\ \bibinfo {author} {\bibfnamefont {P.~A.}\
  \bibnamefont {Maia~Neto}},\ }\bibfield  {title} {\bibinfo {title}
  {{Decoherence via the Dynamical Casimir Effect}},\ }\href
  {https://doi.org/10.1103/PhysRevLett.84.798} {\bibfield  {journal} {\bibinfo
  {journal} {Phys. Rev. Lett.}\ }\textbf {\bibinfo {volume} {84}},\ \bibinfo
  {pages} {798} (\bibinfo {year} {2000})}\BibitemShut {NoStop}%
\bibitem [{\citenamefont {Maia~Neto}\ and\ \citenamefont
  {Dalvit}(2000)}]{MaiaNeto-PRA-2000}%
  \BibitemOpen
  \bibfield  {author} {\bibinfo {author} {\bibfnamefont {P.~A.}\ \bibnamefont
  {Maia~Neto}}\ and\ \bibinfo {author} {\bibfnamefont {D.~A.~R.}\ \bibnamefont
  {Dalvit}},\ }\bibfield  {title} {\bibinfo {title} {Radiation pressure as a
  source of decoherence},\ }\href {https://doi.org/10.1103/PhysRevA.62.042103}
  {\bibfield  {journal} {\bibinfo  {journal} {Phys. Rev. A}\ }\textbf {\bibinfo
  {volume} {62}},\ \bibinfo {pages} {042103} (\bibinfo {year}
  {2000})}\BibitemShut {NoStop}%
\bibitem [{\citenamefont {Sala}\ and\ \citenamefont
  {Tufarelli}(2018)}]{Tufarellu2018}%
  \BibitemOpen
  \bibfield  {author} {\bibinfo {author} {\bibfnamefont {K.}~\bibnamefont
  {Sala}}\ and\ \bibinfo {author} {\bibfnamefont {T.}~\bibnamefont
  {Tufarelli}},\ }\bibfield  {title} {\bibinfo {title} {Exploring corrections
  to the optomechanical hamiltonian},\ }\href
  {https://doi.org/10.1038/s41598-018-26739-0} {\bibfield  {journal} {\bibinfo
  {journal} {Scientific reports}\ }\textbf {\bibinfo {volume} {8}},\ \bibinfo
  {pages} {9157} (\bibinfo {year} {2018})}\BibitemShut {NoStop}%
\bibitem [{\citenamefont {Thompson}\ \emph {et~al.}(2008)\citenamefont
  {Thompson}, \citenamefont {Zwickl}, \citenamefont {Jayich}, \citenamefont
  {Marquardt}, \citenamefont {Girvin},\ and\ \citenamefont
  {Harris}}]{Thompson2008}%
  \BibitemOpen
  \bibfield  {author} {\bibinfo {author} {\bibfnamefont {J.~D.}\ \bibnamefont
  {Thompson}}, \bibinfo {author} {\bibfnamefont {B.~M.}\ \bibnamefont
  {Zwickl}}, \bibinfo {author} {\bibfnamefont {A.~M.}\ \bibnamefont {Jayich}},
  \bibinfo {author} {\bibfnamefont {F.}~\bibnamefont {Marquardt}}, \bibinfo
  {author} {\bibfnamefont {S.~M.}\ \bibnamefont {Girvin}},\ and\ \bibinfo
  {author} {\bibfnamefont {J.~G.~E.}\ \bibnamefont {Harris}},\ }\bibfield
  {title} {\bibinfo {title} {Strong dispersive coupling of a high-finesse
  cavity to a micromechanical membrane},\ }\href
  {https://doi.org/10.1038/nature06715} {\bibfield  {journal} {\bibinfo
  {journal} {Nature}\ }\textbf {\bibinfo {volume} {452}},\ \bibinfo {pages}
  {72} (\bibinfo {year} {2008})}\BibitemShut {NoStop}%
\bibitem [{\citenamefont {Miao}\ \emph {et~al.}(2009)\citenamefont {Miao},
  \citenamefont {Danilishin}, \citenamefont {Corbitt},\ and\ \citenamefont
  {Chen}}]{Miao2009}%
  \BibitemOpen
  \bibfield  {author} {\bibinfo {author} {\bibfnamefont {H.}~\bibnamefont
  {Miao}}, \bibinfo {author} {\bibfnamefont {S.}~\bibnamefont {Danilishin}},
  \bibinfo {author} {\bibfnamefont {T.}~\bibnamefont {Corbitt}},\ and\ \bibinfo
  {author} {\bibfnamefont {Y.}~\bibnamefont {Chen}},\ }\bibfield  {title}
  {\bibinfo {title} {Standard quantum limit for probing mechanical energy
  quantization},\ }\href {https://doi.org/10.1103/PhysRevLett.103.100402}
  {\bibfield  {journal} {\bibinfo  {journal} {Phys. Rev. Lett.}\ }\textbf
  {\bibinfo {volume} {103}},\ \bibinfo {pages} {100402} (\bibinfo {year}
  {2009})}\BibitemShut {NoStop}%
\bibitem [{\citenamefont {Ludwig}\ \emph {et~al.}(2012)\citenamefont {Ludwig},
  \citenamefont {Safavi-Naeini}, \citenamefont {Painter},\ and\ \citenamefont
  {Marquardt}}]{Marquardt2012}%
  \BibitemOpen
  \bibfield  {author} {\bibinfo {author} {\bibfnamefont {M.}~\bibnamefont
  {Ludwig}}, \bibinfo {author} {\bibfnamefont {A.~H.}\ \bibnamefont
  {Safavi-Naeini}}, \bibinfo {author} {\bibfnamefont {O.}~\bibnamefont
  {Painter}},\ and\ \bibinfo {author} {\bibfnamefont {F.}~\bibnamefont
  {Marquardt}},\ }\bibfield  {title} {\bibinfo {title} {Enhanced quantum
  nonlinearities in a two-mode optomechanical system},\ }\href
  {https://doi.org/10.1103/PhysRevLett.109.063601} {\bibfield  {journal}
  {\bibinfo  {journal} {Phys. Rev. Lett.}\ }\textbf {\bibinfo {volume} {109}},\
  \bibinfo {pages} {063601} (\bibinfo {year} {2012})}\BibitemShut {NoStop}%
\bibitem [{\citenamefont {Clerk}\ \emph {et~al.}(2010)\citenamefont {Clerk},
  \citenamefont {Marquardt},\ and\ \citenamefont {Harris}}]{Clerk2010}%
  \BibitemOpen
  \bibfield  {author} {\bibinfo {author} {\bibfnamefont {A.~A.}\ \bibnamefont
  {Clerk}}, \bibinfo {author} {\bibfnamefont {F.}~\bibnamefont {Marquardt}},\
  and\ \bibinfo {author} {\bibfnamefont {J.~G.~E.}\ \bibnamefont {Harris}},\
  }\bibfield  {title} {\bibinfo {title} {Quantum measurement of phonon shot
  noise},\ }\href {https://doi.org/10.1103/PhysRevLett.104.213603} {\bibfield
  {journal} {\bibinfo  {journal} {Phys. Rev. Lett.}\ }\textbf {\bibinfo
  {volume} {104}},\ \bibinfo {pages} {213603} (\bibinfo {year}
  {2010})}\BibitemShut {NoStop}%
\bibitem [{\citenamefont {Bhattacharya}\ \emph {et~al.}(2008)\citenamefont
  {Bhattacharya}, \citenamefont {Uys},\ and\ \citenamefont
  {Meystre}}]{Meystre2008}%
  \BibitemOpen
  \bibfield  {author} {\bibinfo {author} {\bibfnamefont {M.}~\bibnamefont
  {Bhattacharya}}, \bibinfo {author} {\bibfnamefont {H.}~\bibnamefont {Uys}},\
  and\ \bibinfo {author} {\bibfnamefont {P.}~\bibnamefont {Meystre}},\
  }\bibfield  {title} {\bibinfo {title} {Optomechanical trapping and cooling of
  partially reflective mirrors},\ }\href
  {https://doi.org/10.1103/PhysRevA.77.033819} {\bibfield  {journal} {\bibinfo
  {journal} {Phys. Rev. A}\ }\textbf {\bibinfo {volume} {77}},\ \bibinfo
  {pages} {033819} (\bibinfo {year} {2008})}\BibitemShut {NoStop}%
\bibitem [{\citenamefont {Nunnenkamp}\ \emph {et~al.}(2010)\citenamefont
  {Nunnenkamp}, \citenamefont {B\o{}rkje}, \citenamefont {Harris},\ and\
  \citenamefont {Girvin}}]{Girvin2010}%
  \BibitemOpen
  \bibfield  {author} {\bibinfo {author} {\bibfnamefont {A.}~\bibnamefont
  {Nunnenkamp}}, \bibinfo {author} {\bibfnamefont {K.}~\bibnamefont
  {B\o{}rkje}}, \bibinfo {author} {\bibfnamefont {J.~G.~E.}\ \bibnamefont
  {Harris}},\ and\ \bibinfo {author} {\bibfnamefont {S.~M.}\ \bibnamefont
  {Girvin}},\ }\bibfield  {title} {\bibinfo {title} {Cooling and squeezing via
  quadratic optomechanical coupling},\ }\href
  {https://doi.org/10.1103/PhysRevA.82.021806} {\bibfield  {journal} {\bibinfo
  {journal} {Phys. Rev. A}\ }\textbf {\bibinfo {volume} {82}},\ \bibinfo
  {pages} {021806} (\bibinfo {year} {2010})}\BibitemShut {NoStop}%
\bibitem [{\citenamefont {Asjad}\ \emph {et~al.}(2014)\citenamefont {Asjad},
  \citenamefont {Agarwal}, \citenamefont {Kim}, \citenamefont {Tombesi},
  \citenamefont {Giuseppe},\ and\ \citenamefont {Vitali}}]{Vitali2014}%
  \BibitemOpen
  \bibfield  {author} {\bibinfo {author} {\bibfnamefont {M.}~\bibnamefont
  {Asjad}}, \bibinfo {author} {\bibfnamefont {G.~S.}\ \bibnamefont {Agarwal}},
  \bibinfo {author} {\bibfnamefont {M.~S.}\ \bibnamefont {Kim}}, \bibinfo
  {author} {\bibfnamefont {P.}~\bibnamefont {Tombesi}}, \bibinfo {author}
  {\bibfnamefont {G.~D.}\ \bibnamefont {Giuseppe}},\ and\ \bibinfo {author}
  {\bibfnamefont {D.}~\bibnamefont {Vitali}},\ }\bibfield  {title} {\bibinfo
  {title} {Robust stationary mechanical squeezing in a kicked quadratic
  optomechanical system},\ }\href {https://doi.org/10.1103/PhysRevA.89.023849}
  {\bibfield  {journal} {\bibinfo  {journal} {Phys. Rev. A}\ }\textbf {\bibinfo
  {volume} {89}},\ \bibinfo {pages} {023849} (\bibinfo {year}
  {2014})}\BibitemShut {NoStop}%
\bibitem [{\citenamefont {Doolin}\ \emph {et~al.}(2014)\citenamefont {Doolin},
  \citenamefont {Hauer}, \citenamefont {Kim}, \citenamefont {MacDonald},
  \citenamefont {Ramp},\ and\ \citenamefont {Davis}}]{Davis2014}%
  \BibitemOpen
  \bibfield  {author} {\bibinfo {author} {\bibfnamefont {C.}~\bibnamefont
  {Doolin}}, \bibinfo {author} {\bibfnamefont {B.~D.}\ \bibnamefont {Hauer}},
  \bibinfo {author} {\bibfnamefont {P.~H.}\ \bibnamefont {Kim}}, \bibinfo
  {author} {\bibfnamefont {A.~J.~R.}\ \bibnamefont {MacDonald}}, \bibinfo
  {author} {\bibfnamefont {H.}~\bibnamefont {Ramp}},\ and\ \bibinfo {author}
  {\bibfnamefont {J.~P.}\ \bibnamefont {Davis}},\ }\bibfield  {title} {\bibinfo
  {title} {Nonlinear optomechanics in the stationary regime},\ }\href
  {https://doi.org/10.1103/PhysRevA.89.053838} {\bibfield  {journal} {\bibinfo
  {journal} {Phys. Rev. A}\ }\textbf {\bibinfo {volume} {89}},\ \bibinfo
  {pages} {053838} (\bibinfo {year} {2014})}\BibitemShut {NoStop}%
\bibitem [{\citenamefont {Brawley}\ \emph {et~al.}(2016)\citenamefont
  {Brawley}, \citenamefont {Vanner}, \citenamefont {Larsen}, \citenamefont
  {Schmid}, \citenamefont {Boisen},\ and\ \citenamefont {Bowen}}]{Bowen2016}%
  \BibitemOpen
  \bibfield  {author} {\bibinfo {author} {\bibfnamefont {G.~A.}\ \bibnamefont
  {Brawley}}, \bibinfo {author} {\bibfnamefont {M.~R.}\ \bibnamefont {Vanner}},
  \bibinfo {author} {\bibfnamefont {P.~E.}\ \bibnamefont {Larsen}}, \bibinfo
  {author} {\bibfnamefont {S.}~\bibnamefont {Schmid}}, \bibinfo {author}
  {\bibfnamefont {A.}~\bibnamefont {Boisen}},\ and\ \bibinfo {author}
  {\bibfnamefont {W.~P.}\ \bibnamefont {Bowen}},\ }\bibfield  {title} {\bibinfo
  {title} {Nonlinear optomechanical measurement of mechanical motion},\ }\href
  {https://doi.org/10.1038/ncomms10988} {\bibfield  {journal} {\bibinfo
  {journal} {Nat. Commun.}\ }\textbf {\bibinfo {volume} {7}},\ \bibinfo {pages}
  {10988} (\bibinfo {year} {2016})}\BibitemShut {NoStop}%
\bibitem [{\citenamefont {Kaviani}\ \emph {et~al.}(2015)\citenamefont
  {Kaviani}, \citenamefont {Healey}, \citenamefont {Wu}, \citenamefont
  {Ghobadi}, \citenamefont {Hryciw},\ and\ \citenamefont
  {Barclay}}]{Barclay2015}%
  \BibitemOpen
  \bibfield  {author} {\bibinfo {author} {\bibfnamefont {H.}~\bibnamefont
  {Kaviani}}, \bibinfo {author} {\bibfnamefont {C.}~\bibnamefont {Healey}},
  \bibinfo {author} {\bibfnamefont {M.}~\bibnamefont {Wu}}, \bibinfo {author}
  {\bibfnamefont {R.}~\bibnamefont {Ghobadi}}, \bibinfo {author} {\bibfnamefont
  {A.}~\bibnamefont {Hryciw}},\ and\ \bibinfo {author} {\bibfnamefont {P.~E.}\
  \bibnamefont {Barclay}},\ }\bibfield  {title} {\bibinfo {title} {Nonlinear
  optomechanical paddle nanocavities},\ }\href
  {https://doi.org/10.1364/OPTICA.2.000271} {\bibfield  {journal} {\bibinfo
  {journal} {Optica}\ }\textbf {\bibinfo {volume} {2}},\ \bibinfo {pages} {271}
  (\bibinfo {year} {2015})}\BibitemShut {NoStop}%
\bibitem [{\citenamefont {Para\"{\i}so}\ \emph {et~al.}(2015)\citenamefont
  {Para\"{\i}so}, \citenamefont {Kalaee}, \citenamefont {Zang}, \citenamefont
  {Pfeifer}, \citenamefont {Marquardt},\ and\ \citenamefont
  {Painter}}]{Painter2015}%
  \BibitemOpen
  \bibfield  {author} {\bibinfo {author} {\bibfnamefont {T.~K.}\ \bibnamefont
  {Para\"{\i}so}}, \bibinfo {author} {\bibfnamefont {M.}~\bibnamefont
  {Kalaee}}, \bibinfo {author} {\bibfnamefont {L.}~\bibnamefont {Zang}},
  \bibinfo {author} {\bibfnamefont {H.}~\bibnamefont {Pfeifer}}, \bibinfo
  {author} {\bibfnamefont {F.}~\bibnamefont {Marquardt}},\ and\ \bibinfo
  {author} {\bibfnamefont {O.}~\bibnamefont {Painter}},\ }\bibfield  {title}
  {\bibinfo {title} {Position-squared coupling in a tunable photonic crystal
  optomechanical cavity},\ }\href {https://doi.org/10.1103/PhysRevX.5.041024}
  {\bibfield  {journal} {\bibinfo  {journal} {Phys. Rev. X}\ }\textbf {\bibinfo
  {volume} {5}},\ \bibinfo {pages} {041024} (\bibinfo {year}
  {2015})}\BibitemShut {NoStop}%
\bibitem [{\citenamefont {Primo}\ \emph {et~al.}(2023)\citenamefont {Primo},
  \citenamefont {Pinho}, \citenamefont {Benevides}, \citenamefont
  {Gr{\"o}blacher}, \citenamefont {Wiederhecker},\ and\ \citenamefont
  {Alegre}}]{primo2023dissipative}%
  \BibitemOpen
  \bibfield  {author} {\bibinfo {author} {\bibfnamefont {A.~G.}\ \bibnamefont
  {Primo}}, \bibinfo {author} {\bibfnamefont {P.~V.}\ \bibnamefont {Pinho}},
  \bibinfo {author} {\bibfnamefont {R.}~\bibnamefont {Benevides}}, \bibinfo
  {author} {\bibfnamefont {S.}~\bibnamefont {Gr{\"o}blacher}}, \bibinfo
  {author} {\bibfnamefont {G.~S.}\ \bibnamefont {Wiederhecker}},\ and\ \bibinfo
  {author} {\bibfnamefont {T.~P.~M.}\ \bibnamefont {Alegre}},\ }\bibfield
  {title} {\bibinfo {title} {Dissipative optomechanics in high-frequency
  nanomechanical resonators},\ }\href
  {https://doi.org/10.1038/s41467-023-41127-7} {\bibfield  {journal} {\bibinfo
  {journal} {Nature communications}\ }\textbf {\bibinfo {volume} {14}},\
  \bibinfo {pages} {5793} (\bibinfo {year} {2023})}\BibitemShut {NoStop}%
\bibitem [{\citenamefont {Milonni}(1994)}]{milonni-book}%
  \BibitemOpen
  \bibfield  {author} {\bibinfo {author} {\bibfnamefont {P.~W.}\ \bibnamefont
  {Milonni}},\ }\href@noop {} {\emph {\bibinfo {title} {The quantum vacuum: an
  introduction to quantum electrodynamics}}}\ (\bibinfo  {publisher} {Academic
  press New York},\ \bibinfo {year} {1994})\BibitemShut {NoStop}%
\bibitem [{\citenamefont {Bethe}(1947)}]{Bethe-Lamb-1947}%
  \BibitemOpen
  \bibfield  {author} {\bibinfo {author} {\bibfnamefont {H.~A.}\ \bibnamefont
  {Bethe}},\ }\bibfield  {title} {\bibinfo {title} {The electromagnetic shift
  of energy levels},\ }\href {https://doi.org/10.1103/PhysRev.72.339}
  {\bibfield  {journal} {\bibinfo  {journal} {Phys. Rev.}\ }\textbf {\bibinfo
  {volume} {72}},\ \bibinfo {pages} {339} (\bibinfo {year} {1947})}\BibitemShut
  {NoStop}%
\bibitem [{\citenamefont {Ashcroft}\ and\ \citenamefont
  {Mermin}(1976)}]{Ashcroft-book}%
  \BibitemOpen
  \bibfield  {author} {\bibinfo {author} {\bibfnamefont {N.~W.}\ \bibnamefont
  {Ashcroft}}\ and\ \bibinfo {author} {\bibfnamefont {N.~D.}\ \bibnamefont
  {Mermin}},\ }\href@noop {} {\emph {\bibinfo {title} {Solid State Physics}}}\
  (\bibinfo  {publisher} {Saunders College, Philadelphia},\ \bibinfo {year}
  {1976})\BibitemShut {NoStop}%
\bibitem [{\citenamefont {Montalbano}\ \emph {et~al.}(2023)\citenamefont
  {Montalbano}, \citenamefont {Armata}, \citenamefont {Rizzuto},\ and\
  \citenamefont {Passante}}]{Montalbano-2023}%
  \BibitemOpen
  \bibfield  {author} {\bibinfo {author} {\bibfnamefont {F.}~\bibnamefont
  {Montalbano}}, \bibinfo {author} {\bibfnamefont {F.}~\bibnamefont {Armata}},
  \bibinfo {author} {\bibfnamefont {L.}~\bibnamefont {Rizzuto}},\ and\ \bibinfo
  {author} {\bibfnamefont {R.}~\bibnamefont {Passante}},\ }\bibfield  {title}
  {\bibinfo {title} {Spatial correlations of field observables in two
  half-spaces separated by a movable perfect mirror},\ }\href
  {https://doi.org/10.1103/PhysRevD.107.056007} {\bibfield  {journal} {\bibinfo
   {journal} {Phys. Rev. D}\ }\textbf {\bibinfo {volume} {107}},\ \bibinfo
  {pages} {056007} (\bibinfo {year} {2023})}\BibitemShut {NoStop}%
\bibitem [{\citenamefont {Ferreri}\ \emph {et~al.}(2024)\citenamefont
  {Ferreri}, \citenamefont {Bruschi}, \citenamefont {Wilhelm}, \citenamefont
  {Nori},\ and\ \citenamefont {Macr\`{\i}}}]{Ferreri-2024}%
  \BibitemOpen
  \bibfield  {author} {\bibinfo {author} {\bibfnamefont {A.}~\bibnamefont
  {Ferreri}}, \bibinfo {author} {\bibfnamefont {D.~E.}\ \bibnamefont
  {Bruschi}}, \bibinfo {author} {\bibfnamefont {F.~K.}\ \bibnamefont
  {Wilhelm}}, \bibinfo {author} {\bibfnamefont {F.}~\bibnamefont {Nori}},\ and\
  \bibinfo {author} {\bibfnamefont {V.}~\bibnamefont {Macr\`{\i}}},\ }\bibfield
   {title} {\bibinfo {title} {Phonon-photon conversion as mechanism for cooling
  and coherence transfer},\ }\href
  {https://doi.org/10.1103/PhysRevResearch.6.023320} {\bibfield  {journal}
  {\bibinfo  {journal} {Phys. Rev. Res.}\ }\textbf {\bibinfo {volume} {6}},\
  \bibinfo {pages} {023320} (\bibinfo {year} {2024})}\BibitemShut {NoStop}%
\bibitem [{\citenamefont {Butera}(2022)}]{Butera-InfFunc-2022}%
  \BibitemOpen
  \bibfield  {author} {\bibinfo {author} {\bibfnamefont {S.}~\bibnamefont
  {Butera}},\ }\bibfield  {title} {\bibinfo {title} {Influence functional for
  two mirrors interacting via radiation pressure},\ }\href
  {https://doi.org/10.1103/PhysRevD.105.016023} {\bibfield  {journal} {\bibinfo
   {journal} {Phys. Rev. D}\ }\textbf {\bibinfo {volume} {105}},\ \bibinfo
  {pages} {016023} (\bibinfo {year} {2022})}\BibitemShut {NoStop}%
\bibitem [{\citenamefont {Russo}\ \emph {et~al.}(2023)\citenamefont {Russo},
  \citenamefont {Mercurio}, \citenamefont {Mauceri}, \citenamefont {Lo~Franco},
  \citenamefont {Nori}, \citenamefont {Savasta},\ and\ \citenamefont
  {Macr\`{\i}}}]{Savasta-Hopping-2023}%
  \BibitemOpen
  \bibfield  {author} {\bibinfo {author} {\bibfnamefont {E.}~\bibnamefont
  {Russo}}, \bibinfo {author} {\bibfnamefont {A.}~\bibnamefont {Mercurio}},
  \bibinfo {author} {\bibfnamefont {F.}~\bibnamefont {Mauceri}}, \bibinfo
  {author} {\bibfnamefont {R.}~\bibnamefont {Lo~Franco}}, \bibinfo {author}
  {\bibfnamefont {F.}~\bibnamefont {Nori}}, \bibinfo {author} {\bibfnamefont
  {S.}~\bibnamefont {Savasta}},\ and\ \bibinfo {author} {\bibfnamefont
  {V.}~\bibnamefont {Macr\`{\i}}},\ }\bibfield  {title} {\bibinfo {title}
  {Optomechanical two-photon hopping},\ }\href
  {https://doi.org/10.1103/PhysRevResearch.5.013221} {\bibfield  {journal}
  {\bibinfo  {journal} {Phys. Rev. Res.}\ }\textbf {\bibinfo {volume} {5}},\
  \bibinfo {pages} {013221} (\bibinfo {year} {2023})}\BibitemShut {NoStop}%
\bibitem [{\citenamefont {Mercurio}\ \emph {et~al.}(2025)\citenamefont
  {Mercurio}, \citenamefont {Russo}, \citenamefont {Mauceri}, \citenamefont
  {Savasta}, \citenamefont {Nori}, \citenamefont {Macr\`{\i}},\ and\
  \citenamefont {Lo~Franco}}]{Savasta-SciPost-2025}%
  \BibitemOpen
  \bibfield  {author} {\bibinfo {author} {\bibfnamefont {A.}~\bibnamefont
  {Mercurio}}, \bibinfo {author} {\bibfnamefont {E.}~\bibnamefont {Russo}},
  \bibinfo {author} {\bibfnamefont {F.}~\bibnamefont {Mauceri}}, \bibinfo
  {author} {\bibfnamefont {S.}~\bibnamefont {Savasta}}, \bibinfo {author}
  {\bibfnamefont {F.}~\bibnamefont {Nori}}, \bibinfo {author} {\bibfnamefont
  {V.}~\bibnamefont {Macr\`{\i}}},\ and\ \bibinfo {author} {\bibfnamefont
  {R.}~\bibnamefont {Lo~Franco}},\ }\bibfield  {title} {\bibinfo {title}
  {Bilateral photon emission from a vibrating mirror and multiphoton
  entanglement generation},\ }\href
  {https://doi.org/10.21468/SciPostPhys.18.2.067} {\bibfield  {journal}
  {\bibinfo  {journal} {SciPost Phys.}\ }\textbf {\bibinfo {volume} {18}},\
  \bibinfo {pages} {067} (\bibinfo {year} {2025})}\BibitemShut {NoStop}%
\bibitem [{\citenamefont {Schneiter}\ \emph {et~al.}(2020)\citenamefont
  {Schneiter}, \citenamefont {Qvarfort}, \citenamefont {Serafini},
  \citenamefont {Xuereb}, \citenamefont {Braun}, \citenamefont {R\"atzel},\
  and\ \citenamefont {Bruschi}}]{Schneiter-PRA-2020}%
  \BibitemOpen
  \bibfield  {author} {\bibinfo {author} {\bibfnamefont {F.}~\bibnamefont
  {Schneiter}}, \bibinfo {author} {\bibfnamefont {S.}~\bibnamefont {Qvarfort}},
  \bibinfo {author} {\bibfnamefont {A.}~\bibnamefont {Serafini}}, \bibinfo
  {author} {\bibfnamefont {A.}~\bibnamefont {Xuereb}}, \bibinfo {author}
  {\bibfnamefont {D.}~\bibnamefont {Braun}}, \bibinfo {author} {\bibfnamefont
  {D.}~\bibnamefont {R\"atzel}},\ and\ \bibinfo {author} {\bibfnamefont
  {D.~E.}\ \bibnamefont {Bruschi}},\ }\bibfield  {title} {\bibinfo {title}
  {Optimal estimation with quantum optomechanical systems in the nonlinear
  regime},\ }\href {https://doi.org/10.1103/PhysRevA.101.033834} {\bibfield
  {journal} {\bibinfo  {journal} {Phys. Rev. A}\ }\textbf {\bibinfo {volume}
  {101}},\ \bibinfo {pages} {033834} (\bibinfo {year} {2020})}\BibitemShut
  {NoStop}%
\end{thebibliography}%

\end{document}